\documentclass[12pt,journal,draftcls,onecolumn,a4paper]{IEEEtran}
\usepackage{epsfig}
\usepackage[latin1]{inputenc}
\usepackage{cite,amsfonts}
\usepackage{longtable}
\usepackage{bm,amsmath,amssymb,psfrag,dsfont,amsthm,graphicx}

\usepackage{anysize}  % CREO QUE NECESISTAS ESTE PAQUETE PARA LOS MARGENES
\marginsize{1.8cm}{1.8cm}{1.6cm}{1.6cm} % iz, dere, arri, aba
\newtheorem{lemma}{Lemma}

\ifCLASSINFOpdf
\else
\fi
\begin{document}
\title{Optimization of Radio and Computational Resources for Energy Efficiency in Latency-Constrained Application Offloading}

\author{Olga Muñoz, Antonio Pascual-Iserte, Josep Vidal\\ \small{Dept. of Signal Theory and Communications}\\ Universitat Politècnica de Catalunya (UPC), Barcelona, Spain\\ Emails: \{olga.munoz, antonio.pascual, josep.vidal\}@upc.edu.% <-this % stops a space
\thanks{Address: c/ Jordi Girona 1-3, mòdul D5 - campus nord UPC, 08034 Barcelona, Spain.}
\thanks{The research leading to these results has received funding from the European Commission (FP7) through project TROPIC INFSO-ICT-318784 and network of excellence NEWCOM\# (grant agreement 318306), from the Spanish Ministry of Economy and Competitiveness (Ministerio
de Economía y Competitividad) through projects TEC2013-41315-R
(DISNET) and TEC2011-29006-C03-02 (GRE3N-LINK-MAC), and from the Catalan Government through grant 2014 SGR 60 (AGAUR).}
\thanks{Part of this work has been presented at Future Network \& Mobile Summit (FUNEMS), Lisbon (Portugal), July 2013.}}


\markboth{IEEE Transactions on Vehicular Technology, submitted October 2014}%
{Shell \MakeLowercase{\textit{et al.}}: Bare Demo of IEEEtran.cls for Journals}



\maketitle


\begin{abstract}
Providing femto-access points (FAPs) with
computational capabilities will allow (either total or partial)
offloading of highly demanding applications from smart-phones to the
so called femto-cloud. Such offloading promises to be beneficial in
terms of battery saving at the mobile terminal (MT) and/or in latency
reduction in the execution of applications. However, for this promise to become a reality, the energy and/or the time
required for the communication process are compensated by the energy
and/or the time savings that result from the remote computation at the FAPs. For this
problem, we provide in this paper a framework for the joint optimization of the radio and computational resource usage exploiting the tradeoff between energy consumption and latency. Multiple antennas are assumed to be available at the MT and the serving FAP. As a result of the optimization, the optimal communication strategy (e.g., transmission power, rate, precoder) is obtained, as well as the optimal
distribution of the computational load between the handset and the serving FAP. This paper also establishes the conditions under which total or no offloading are optimal, determines which is the minimum affordable latency in the execution of the application, and analyzes as a particular case the minimization of the total consumed energy without latency constraints.
\end{abstract}

\begin{IEEEkeywords}
Femto-cloud, offloading, battery saving,
energy-latency trade-off, energy efficiency, multi-input multi-output (MIMO).
\end{IEEEkeywords}

% \fi
\IEEEpeerreviewmaketitle

\newpage

\section{Introduction}
Cloud computing is becoming a key flexible and cost-effective tool to allow mobile terminals (MTs) to have access to much larger computational and storage resources than those available in typical user equipments. Furthermore, reducing the computational effort of the MTs may help to extend the lifetime of the batteries, which is currently an important limitation of user devices such as smart-phones. At the same time, an exponential growth of femto deployments is expected \cite{chandrasekhar:08,luening:09} due, in part, to the fact that spatial proximity between the handset and the serving femto access point (FAP) enables successful communication with high rates and reduced power. In this context, femtocell deployments can be seen as an opportunity to offer low-cost solutions for cloud services by equipping the FAPs with some amount of computational and storage capabilities. By exploiting the virtualization and distribution paradigms employed in cloud services, very demanding applications for MTs in terms of computation, storage, and latency could be distributed over cooperative FAPs. This idea was already presented in \cite{zhu:11} under the concept of media-edge cloud for multimedia computing.

The challenges of supporting mobile cloud computing applications include, but are not limited to, the offloading decision criteria, admission control, cell association, power control, and resource allocation \cite{lei:13}. Most of the work done so far corresponds to the management aspects, the experimental evaluation of the energy saving associated to the offloading, and/or the definition of an offloading criterion that takes into account the energy cost of the radio interface (e.g., 3G or WiFi) but without optimizing the energy cost of the data transfer according to the current channel conditions \cite{gkatzikis:13,miettinen:10,kosta:12,zhu:11,cuervo:10,kumar:10,kumar:13,lagerspetz:11,kovachev:12}. Notice, however, that depending not only on the application but also on the current channel conditions, the best strategy as far as the offloading process is concerned may be different. This radio-cloud interaction is addressed in \cite{zhang:13}, by considering the Gilbert-Elliott channel model for the wireless transmission. While that model may provide some hints about the impact of the quality of the wireless link on the transmission rate and the offloading decision, it does not consider the optimization of the precoding strategy for the offloading when multiple antennas are available (i.e., multi-input multi-output (MIMO) channels) or the inclusion of practical constraints such as the maximum transmission power available at both the MT and the serving FAP. On the other hand, this model includes the energy cost when the MT is transmitting but not when the MT is receiving, and so the downlink (DL) is not considered in the analysis carried out in \cite{zhang:13}.

In this paper, we provide a framework for the joint optimization of the computational and radio resources usage in the described scenario assuming that multiple antennas are available simultaneously at the MT and the FAP. As a result of the optimization, the optimal transmission strategy will be obtained (including the transmission power, the precoder, and the rate for transferring the data in both uplink (UL) and DL), as well as the optimal distribution of the computational load between the MT and the FAP. As in \cite{miettinen:10,kosta:12,zhu:11,kovachev:12,zhang:13}, energy consumption and also total execution time are the key performance indicators considered for the optimization. However, our work presents some differences w.r.t. previous works. Firstly, instead of considering that the application is run either totally at the cloud or totally at the MT, we include as an optimization variable the amount of data to be processed at each side and show under what conditions parallelizing the processing is optimum. Secondly, different from previous works, our approach allows adapting the transmission strategy to the current channel as perceived by the MT in the DL, and by the serving FAP in the UL, and includes practical aspects such as the maximum radiated powers and the maximum rate supported by the system. More importantly, our analysis provides the optimum transmission strategy for the offloading in a MIMO set up, which goes beyond the optimal MIMO strategy when considering a stand-alone communication problem where the objective is only the maximization of the mutual information or the minimization of the transmission power \cite{palomar:03}. This aspect represents a step forward w.r.t. other works in the literature related to offloading such as \cite{zhang:13}. Finally, our analysis includes the derivation of the conditions under which total or no offloading are optimum, the minimum energy required to execute an application with no latency constraints, and the minimum required time budget.

Our paper is a generalization of the results presented by the same authors in the conference paper \cite{munoz:13}. The main novel technical contributions w.r.t. that paper are:
\begin{itemize}
\item This paper derives the solution of the general problem and presents results for the case of transmitting through multiple eigenmodes of MIMO channels, whereas in \cite{munoz:13} only the particular cases of single-input single-output (SISO), multi-input single-output (MISO), and single-input multi-output (SIMO) channels were addressed.

\item An in-depth theoretical analysis of the functions describing the inherent tradeoff between the latency and the energy spent in the communication is derived, whereas in \cite{munoz:13} only a numerical analysis by means of simulations was provided.

\item Partial closed-form expressions of some key figures of the problem (communication energy, rate, etc.) and a simple one-dimensional convex numerical optimization technique are provided for the resource allocation problem, whereas in \cite{munoz:13} only a multi-dimensional numerical method with high complexity was proposed to solve the problem.

\item This paper analyzes in detail some particular cases derived from the general problem that were not presented in \cite{munoz:13}. These derivations include the optimality of the non-offloading and total offloading approaches, the minimum affordable latency, and the minimum required energy with no latency constraints.
\end{itemize}

We would like to emphasize that this paper focuses on the theoretical radio-cloud interaction of application offloading. Of course, other business and economic aspects could play a fundamental role in the exploitation of this kind of scenarios (see \cite{zhang:10} for a reference describing the business model of cloud computing, or \cite{wang:13} for cloud pricing structures including computing, storage, and network prices). For example, if the application is offloaded to a FAP owned by the user running the application, only technical criteria may be considered when taking the offloading decision. On the other hand, in a ``pay as you go" cloud computing model (i.e., if the user has to pay for the remote execution), the decision could be not to offload the application even if this would be advisable from a technical point of view in terms of energy and/or latency. These economic aspects are, however, beyond the scope of this paper.

It is also important to remark that the analysis carried out in this paper is applicable both to a single user system and to a multiuser system where a set of resources (i.e., bandwidth and CPU rate) have been already pre-allocated (i.e., reserved) to each user. In this framework, we aim to optimize the energy-latency trade-off from the point of view of the MTs to provide insights into how to do an efficient use of the available resources. Due to the lack of space, combining multiuser scheduling with the energy-latency trade-off optimization described here will be considered for future research (some preliminary results by the authors of this paper can be found in \cite{molina:14,munoz:14}).

The rest of the paper is organized as follows. A description of the different kinds of applications and the computational models is provided in Section \ref{sec:computational_model}. Section \ref{sec:scenario_problem} defines the offloading problem and describes the
reference scenario. Section \ref{sec:tradeoff_comm} formulates the adopted power consumption models and the
trade-off between energy and latency in the MIMO wireless
communication link connecting the MT and the FAP. Such trade-off is
exploited in Section \ref{sec:joint_optim_radio_comput} to present a
method to obtain the optimal offloading strategy.
A number of particular cases are analyzed in detail in Section
\ref{sec:particular_cases}. Finally, some simulations results and
conclusions are provided in Sections \ref{sec:simulations} and
\ref{sec:conclusions}, respectively.

\section{Types of Applications and Computational Models}\label{sec:computational_model}
There are a significant number of applications that can fit the ``cloud-service" model. Depending on the type of application, the resource management may need to be tackled in a different way. A possible classification of applications corresponds to the following three major groups:

\begin{enumerate}

\item \emph{Data partitioned oriented applications}. In this type of applications the amount of data to be processed is known beforehand and the execution can be parallelized into processes. Each process takes care of a portion of the total amount of data. An example of this type of applications is a face detection application running over a set of images saved on the user's phone or downloaded from the Internet that counts the number of faces in each picture and computes, for each detected face, simple metrics such as the distance between eyes \cite{barbera:14}. Other examples are a virus scan application, where a set of files are checked to detect possible virus; or a gzip compression application, where a set of files are compressed. Photosynth (http://www.photosynth.net/), a software application that analyzes digital photographs and generates a three-dimensional (3-D) model of the photos after performing image conversion, feature extraction, image matching, and reconstruction, is another example of suitable cloud computing application \cite{zhu:11} that can be classified within this group. Note that in any case, a load balancer divides the whole set of files (or images) into several subsets that are processed in parallel.

\item \emph{Code partitioned oriented applications}. The second type of applications corresponds to applications that can be divided into several methods. Some of the methods can be parallelized; others need to be sequential as the output of some of these methods are the inputs to other ones. This type of applications have been considered in \cite{cuervo:10}. In that paper, the execution dependencies within the program are modeled at a high level using a \emph{call graph}. Assuming that the quantity of input data for each method is known, in addition to the energy and runtime required by the module depending on whether it is running locally or at the cloud, \cite{cuervo:10} obtains the optimal partitioning strategy that minimizes the energy consumed by the smart-phone. Such optimum partitioning is computed before the actual execution starts.

\item \emph{Continuous execution (i.e., real time) applications}. This type of applications includes applications where it is not known beforehand for how long the application is going to be run. Gaming and other interactive applications belong to this group (see as an example the reference to Cloud Mobile Gaming (CMG) in \cite{wang:13}). Note that this type of applications may have different requirements than the previous ones, in the same way as real-time and best-effort traffics have different requirements in stand-alone communication problems. An example of how to deal with this kind of applications can be found in \cite{molina:14}.

\end{enumerate}

In this paper we focus on the first type of applications, i.e., on \emph{data partitioned oriented applications}. Therefore, we will assume that the amount of data to be processed is known before starting the execution and that such execution can be parallelized. The application can be abstracted as a profile with three parameters: (i) the size of the data set $S_{\mbox{\scriptsize{app}}}$ (i.e., the number of data bits to be processed by the application), (ii) the completion deadline $L_{\max}$ (i.e., the maximum value of the delay before which the execution of the application should be completed), and (iii) the output data size (i.e., the number of data bits generated by the execution of the application). In \cite{zhang:13}, the first two parameters were considered for the abstraction.

We evaluate here analytically the impact of the latency requirement on the energy cost and optimize the physical layer parameters (e.g., transmission rate, power, precoder) for an optimum energy-latency trade-off in a complementary way to \cite{barbera:14}, that optimizes the architecture (but not the physical layer transmission) to reduce energy cost without considering latency.

For the analytical developments in this paper, it will be assumed that the data can be partitioned into subsets of any size, despite in practice only some partitions may be possible (for instance, if we are compressing a set of files, the possible partitions depend on the individual sizes of the files to be compressed). That means that in a practical implementation, the optimal solution should be further quantized. In this sense, we claim that the results that we provide in this paper can be understood as a benchmark or upper-bound of the performance of any realistic offloading strategy. Another issue to take into account is the number of CPU cycles needed to complete the job. Following \cite{miettinen:10} and \cite{lorch:01}, \cite{zhang:13} models the number of cycles $N_c$ required to complete the execution of an application with a probability $p$ close to 1, as the product between the number of input bits and a factor that depends on the probability $p$ and the computation complexity of the algorithm. In our paper, we will also model the number of CPU cycles as the number of input bits multiplied by a factor that measures the required CPU cycles per input bit. Meaningful values for the number of CPU cycles per bit obtained from measurements when running real applications can be found in \cite{miettinen:10}.

\section{Description of the Scenario and Problem
Statement}\label{sec:scenario_problem}

\begin{figure}[t]
    \begin{center}
        \includegraphics[width=11cm]{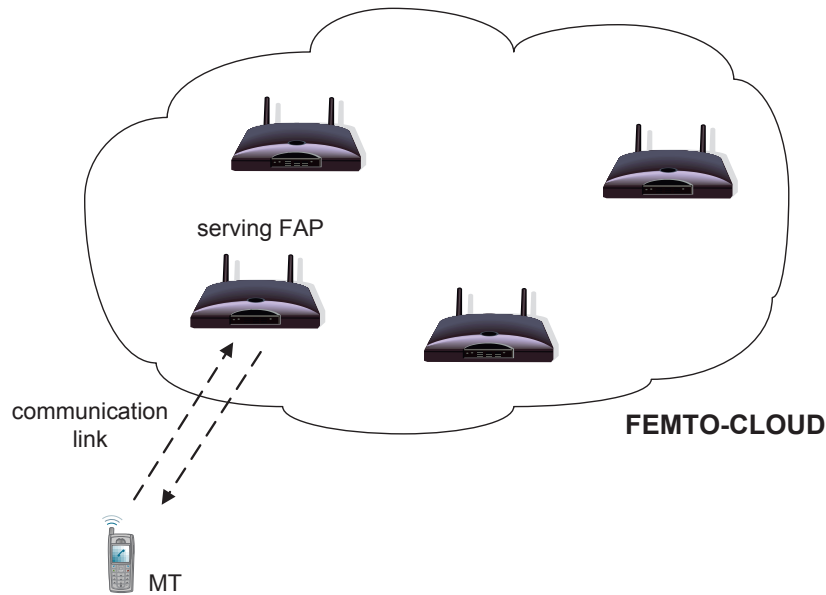}
    \end{center}
\vspace{-0.7cm}
    \caption{Example of a femto-cloud and a MT connected to a serving FAP.}
    \label{fig:scenario}
\end{figure}

We consider a set of FAPs endowed with some storage and
computational capabilities. This set of FAPs forms a femto-cloud, as
shown in Fig. \ref{fig:scenario}. In the most general setup, the application could run in parallel in a
distributed way at the MT, the FAPs in the femto-cloud, or even in
computation entities belonging to other external clouds. In this scenario we
focus on a given MT within the radio range of its serving FAP. We
assume that the user wants to launch an application and it has to be
decided where this application should be executed, namely (i)
totally at the MT, (ii) totally at the femto-cloud, or (iii)
partially at the MT and the femto-cloud (partial offloading). In the
last case, the amount of data to be processed at the MT and the
femto-cloud must be decided as well. When taking the decision,
several aspects should be considered, such as a limited time budget (formulated in terms of a maximum allowed
latency), the total number of bits to be processed, the computational
capabilities of the MT and the femto-cloud, the channel state, and
the energy consumption.

Under the goal of obtaining meaningful insights into the role of the different parameters when evaluating the benefits from the offloading, let us consider in this paper a simple case in the sense that the only element in the
cloud allowed to execute the offloaded processes is the serving FAP.
As far as the application is concerned, we will assume that the only
possible parallelization is between the MT and the serving FAP when
partial offloading is carried out.

The wireless communication channel between the MT and the serving
FAP constitutes the link through which the MT and the FAP exchange data. In case that (partial) offloading is decided, the MT
will send through such link the data to be processed by the FAP and,
once the remote execution is completed, the resulting data will be sent
back from the FAP to the MT. Obviously, the quality of such wireless
channel has a direct impact on the system performance and the
decision to be taken concerning the offloading of the application. Different from \cite{munoz:13}, in this
paper we consider the most general case of having
multiple antennas simultaneously at the MT and the serving FAP,
i.e., a MIMO channel. We will focus on almost static scenarios in the sense that the channel does not change within the maximum latency constraint of the application. This is a reasonable assumption as we are considering that each user is within the range of his/her serving FAP, typically located in indoor scenarios such as homes or offices. Furthermore, due to the low mobility, we assume that the channel is known at both the receiver and transmitter side, through proper feedback. We leave for future research the extension of the proposed techniques to the cases of unknown and/or time-varying channels. In the case that the users have a mobility such that the previous assumption is not valid, the algorithms and strategies that are presented in this paper should be adapted and extended to take this fact into account. Although this falls out of the scope of the paper, in Section \ref{sec:conclusions} devoted to the conclusions and future work, there are some general guidelines and ideas to extend the proposed strategies to the case of time-varying channels.

We focus the attention only on the MTs as far as the energy
consumption is concerned. This is based on the fact that handsets
are battery driven and, therefore, constraining or optimizing their
energy consumption will help to enlarge their lifetimes. Note also that FAPs are usually connected to the electric power grid and, therefore, their lifetime is not limited by the energy consumption. According to this, in this paper we will only formulate the energy for the MTs and will
not include the energy spent by the FAPs. Anyway, if the energy consumption of the FAPs is to be considered as well, this could be done by introducing it into the corresponding power consumption models that will be presented in the following sections in this paper.

Finally, and as it has been already explained in the introduction of this paper, we would like to emphasize again that, although FAPs are multiuser in nature, in this work we have only considered the case of a single-user system or a multiuser scenario where each user has available a certain bandwidth and processor rate and the tradeoff between energy and latency is optimized on a per-user basis. The generalization would imply including in the optimization the distribution of communication bandwidth and processor rate among users as well. Note, however, that this is an extremely complicated task which requires defining a meaningful energy-latency trade-off for all users by introducing, for example, Pareto-optimality concepts. Some preliminary results have been presented by the same authors of this paper in \cite{munoz:14,molina:14}, where the multiuser allocation problem and the optimization of the tradeoff between energy and latency are addressed in a suboptimum way and for concrete scenarios. It is left for future research the analysis of how to solve the general multiuser case in an optimum manner.

\section{Trade-off between Latency and Energy in the Wireless
Transmission}\label{sec:tradeoff_comm}

\subsection{Energy Consumption Model for the MT}
The communication strategy adopted in the physical (PHY) layer will have an
impact on the total energy consumption. In order to optimize the
balance between the energy spent for communication and for computation, under a maximum latency constraint
imposed by the application, we need first of all to provide appropriate
models for the energy consumption associated to the communication. As explained before, the energy spent by the FAPs will not be considered explicitly in this paper.

In modern communications systems, such as LTE, the receiver informs the transmitter about the maximum modulation and coding scheme (MCS) supported \cite{sesia:09}. This MCS translates directly into the achievable rate within the reported bandwidth, which depends on the specific channel conditions as well as on the transmission power. Furthermore, UL power control is supported in LTE systems. Given this, for a certain channel state, the rate supported in the UL may be greater at the expense of increasing the transmission power of the MT and, therefore, its energy consumption. Besides, as a greater MCS increases the encoding and decoding complexity, a greater power supply at the MT may be required. According to this, the purpose of this section is to provide an energy model for the MT relating the power consumption, the radiated power, and the rate, for both the UL and DL transmissions. Of course, this model will have an impact on the offloading optimization, as will be shown later in Section \ref{sec:joint_optim_radio_comput}.

We emphasize that these models (and also the offloading optimization problem in Section \ref{sec:joint_optim_radio_comput}) could be generalized to encompass
also the energy spent by the FAPs when considered appropriate.

\subsubsection{UL Transmission (MT Acting as Transmitter)}
Although there are already some preliminary works covering the power consumption modeling of a transmitter, the truth is that there is no general model universally accepted yet. Under this circumstance, a model that is gaining acceptance is the one provided by the European project EARTH \cite{earth_d23}. Although the scope of this project was focused on the analysis of the energy consumption of the base stations, it should be indicated that the obtained relationships among the variables involved in such model are also valid for the case of the MTs by adjusting properly the parameters appearing in the model. This model is presented in the following.

When the MT transmits through the UL, the radio frequency
(RF) power consumption at the transmitter depends on the radiated
power $p_{\mbox{\scriptsize{tx}}}$, while the power consumed by the
transmitter baseband (BB) processing circuits is affected by the
turbo encoding whose complexity depends on the UL data rate
$r_{\mbox{\tiny{UL}}}$ (defined as the quotient between the bits
transmitted in the UL ($s_{\mbox{\tiny{UL}}}$) and the time
dedicated to the UL transmission ($t_{\mbox{\tiny{UL}}}$), i.e.,
$r_{\mbox{\tiny{UL}}}=\frac{s_{\mbox{\tiny{UL}}}}{t_{\mbox{\tiny{UL}}}}$).
In addition to that, a baseline power is consumed just for having the transmission circuitry switched on. According to the practical measurements
provided in \cite{jensen:12} for a LTE-MT dongle, the UL power consumption $p_{\mbox{\scriptsize{UL}}}$ is greatly affected by the radiated power $p_{\mbox{\scriptsize{tx}}}$ while its dependence w.r.t. $r_{\mbox{\tiny{UL}}}$ due to the encoding is negligible. Based on
these observations, we will adopt the following model for the MT power
consumption in UL:
\begin{equation}
p_{\mbox{\tiny{UL}}}\backsimeq
k_{\mbox{\scriptsize{tx}},1}+k_{\mbox{\scriptsize{tx}},2}p_{\mbox{\scriptsize{tx}}}.\label{eq:ul_power_model}
\end{equation}
In the previous expression, $k_{\mbox{\scriptsize{tx}},1}$ represents the extra power consumption for having the RF and BB transmission circuitries switched on and $k_{\mbox{\scriptsize{tx}},2}$ measures the linear increase of the transmitter power consumption with the radiated power. In the previous model $k_{\mbox{\scriptsize{tx}},2}$ is a scale parameter with no units, whereas $k_{\mbox{\scriptsize{tx}},1}$ has W as units. The expression in (\ref{eq:ul_power_model}) is the one recommended by the EARTH project \cite{earth_d23}. It is important to remark that $p_{\mbox{\tiny{UL}}}$ will depend implicitly on the UL transmission rate $r_{\mbox{\tiny{UL}}}$ through $p_{\mbox{\scriptsize{tx}}}$, as the radiated power will have an impact on the UL signal to noise ratio and, therefore, on the supported UL data rate.

The numerical values of the parameters in the previous  model should be adjusted taking experimental measurements of the energy consumption for a MT. In that sense, \cite{jensen:12} describes an experiment thanks to which a set of real measurements have been obtained. Based on such measurements, it is possible to calculate the numerical values of the model parameters through numerical regressions and to assert that the models obtained in the EARTH project are also valid for the MTs. These numerical values based on the measurements shown in \cite{jensen:12} will be presented when describing some simulations later in this Section and also in Section \ref{sec:simulations}.

\subsubsection{DL Transmission (MT Acting as Receiver)}
When the MT receives through the DL, the RF power
consumption at the receiver may change with the DL received power
level $p_{\mbox{\scriptsize{rx}}}$ (due to the adjustment of the
programmable gain amplifier to adapt the signal level), while the
complexity and, thus, the power spent by the receiver BB processing
circuits, increases linearly with the DL data rate
$r_{\mbox{\tiny{DL}}}$ \cite{cui:03}. The DL rate is defined as the quotient between the bits transmitted in the DL ($s_{\mbox{\tiny{DL}}}$) and the time dedicated to the DL transmission ($t_{\mbox{\tiny{DL}}}$), i.e., $r_{\mbox{\tiny{DL}}}=\frac{s_{\mbox{\tiny{DL}}}}{t_{\mbox{\tiny{DL}}}}$. Finally, a baseline power is also consumed
just for having the reception chain switched on.
The measurements provided in \cite{jensen:12} show that the variation of the DL power consumption $p_{\mbox{\scriptsize{DL}}}$ w.r.t. the DL received
power $p_{\mbox{\scriptsize{rx}}}$ is negligible. Based on these observations, we will use
the following model for the MT power consumption in DL:
\begin{equation}
p_{\mbox{\tiny{DL}}}\backsimeq
k_{\mbox{\scriptsize{rx}},1}+k_{\mbox{\scriptsize{rx}},2}r_{\mbox{\tiny{DL}}}.\label{eq:dl_power_model}
\end{equation}
In the previous expression, $k_{\mbox{\scriptsize{rx}},1}$ represents the extra power consumption for having the reception circuitry switched on and $k_{\mbox{\scriptsize{rx}},2}$ measures the increase of the power consumption with the decoding rate. In the previous model, the parameters $k_{\mbox{\scriptsize{rx}},1}$ and $k_{\mbox{\scriptsize{rx}},2}$ have W and W/bps as units, respectively.

As in the UL case, the numerical values of the parameters $k_{\mbox{\scriptsize{rx}},1}$ and $k_{\mbox{\scriptsize{rx}},2}$ that have been used in the simulations and in the rest of this paper are based on the the measurements provided in \cite{jensen:12}.

\subsection{Trade-off between Latency and Energy in the UL Transmission}

\subsubsection{Computation of the Minimum Energy}
Let us assume a MT with $n_{\mbox{\tiny{MT}}}$ antennas transmitting through the UL a
vector of signals $\mathbf{x}\in\mathds{C}^{n_{\mbox{\tiny{MT}}}\times 1}$. We
define the power transmit covariance matrix as
$\widetilde{\mathbf{Q}}=\mathbb{E}\left[\mathbf{x}\mathbf{x}^H\right]$. According to
(\ref{eq:ul_power_model}), the energy spent by the MT in the UL
transmission can be expressed as
\begin{equation}
k_{\mbox{\scriptsize{tx}},1}t_{\mbox{\tiny{UL}}}+k_{\mbox{\scriptsize{tx}},2}t_{\mbox{\tiny{UL}}}\mbox{Tr}(\widetilde{\mathbf{Q}}),
\end{equation}
where $t_{\mbox{\tiny{UL}}}$ is the time spent by the MT to send $s_{\mbox{\tiny{UL}}}$ information bits.

For any value of $t_{\mbox{\tiny{UL}}}$ and $s_{\mbox{\tiny{UL}}}$, the minimum energy consumed by the MT in the UL transmission, denoted in what follows by $e_{\mbox{\tiny{UL}}}(t_{\mbox{\tiny{UL}}},s_{\mbox{\tiny{UL}}})$, is obtained as the minimum value of the objective function in the following optimization problem:
\begin{equation}
\begin{array}
[c]{rl}
\operatorname*{minimize}\limits_{\widetilde{\mathbf{Q}}} & k_{\mbox{\scriptsize{tx}},1}t_{\mbox{\tiny{UL}}}+k_{\mbox{\scriptsize{tx}},2}t_{\mbox{\tiny{UL}}}\mbox{Tr}(\widetilde{\mathbf{Q}})\\
\operatorname*{subject\,\,to} &
\!\!\begin{array}
[t]{l} C1: s_{\mbox{\tiny{UL}}}\leq
W_{\mbox{\tiny{UL}}}t_{\mbox{\tiny{UL}}}\log_2\left|\mathbf{I}+\mathbf{H}\widetilde{\mathbf{Q}}\mathbf{H}^H\right|,
\\ C2:
\widetilde{\mathbf{Q}}\succeq\mathbf{0}.
\end{array}
\end{array}
\label{eq:energy_latency_power_ul}
\end{equation}

The solution to this problem is well known (see \cite{palomar:03} and \cite{raleigh:98}) and summarized as follows. Let us consider the channel eigendecomposition
$\mathbf{H}^H\mathbf{H}=\mathbf{U}\mathbf{\Lambda}\mathbf{U}^H$,
where $\mathbf{\Lambda}\in\mathds{R}^{n_{\mbox{\tiny{MT}}}\times
n_{\mbox{\tiny{MT}}}}$ is a diagonal matrix with the eigenvalues
$\lambda_i\geq 0$ ($i=1,\ldots,n_{\mbox{\tiny{MT}}}$) in decreasing
order and $\mathbf{U}\in\mathds{C}^{n_{\mbox{\tiny{MT}}}\times
n_{\mbox{\tiny{MT}}}}$ is the unitary matrix whose columns are the
corresponding unit-norm eigenvectors. To minimize the energy consumption in the UL, the UL transmission needs to be done through the channel eigenmodes, applying a power water-filling over them \cite{raleigh:98,palomar:03}, i.e.,
\begin{equation}
\widetilde{\mathbf{Q}}^{\star}=\mathbf{U}\mathbf{P}\mathbf{U}^H,\hspace{0.5cm}\mathbf{P}=\mbox{diag}(\{p_i\}_
{i=1}^{n_{\mbox{\tiny{MT}}}}),\hspace{0.5cm}p_i=\left(c(t_{\mbox{\tiny{UL}}},s_{\mbox{\tiny{UL}}})-\frac{1}{\lambda_i}\right)^{+},
\end{equation}
where $(x)^{+}=\max\{0,x\}$ and $c(t_{\mbox{\tiny{UL}}},s_{\mbox{\tiny{UL}}})$ is a constant
calculated to satisfy constraint C1 in problem
(\ref{eq:energy_latency_power_ul}) with equality. Note that such water-level $c(t_{\mbox{\tiny{UL}}},s_{\mbox{\tiny{UL}}})$ is a function of $t_{\mbox{\tiny{UL}}}$ and $s_{\mbox{\tiny{UL}}}$. Therefore, the number of active eigenmodes, i.e., the number of eigenmodes for which $c(t_{\mbox{\tiny{UL}}},s_{\mbox{\tiny{UL}}})>\frac{1}{\lambda_i}$, will be a function of $t_{\mbox{\tiny{UL}}}$ and $s_{\mbox{\tiny{UL}}}$ as well. Let us, in the following, denote such number of active eigenmodes by $K(t_{\mbox{\tiny{UL}}},s_{\mbox{\tiny{UL}}})$.

Based on the above, for given values of $t_{\mbox{\tiny{UL}}}$ and $s_{\mbox{\tiny{UL}}}$, the minimum energy consumption required for the UL transmission is given by
\begin{equation}
e_{\mbox{\tiny{UL}}}(t_{\mbox{\tiny{UL}}},s_{\mbox{\tiny{UL}}})=k_{\mbox{\scriptsize{tx}},1}t_{\mbox{\tiny{UL}}}+k_{\mbox{\scriptsize{tx}},2}t_{\mbox{\tiny{UL}}}\sum_{i=1}^{K(t_{\mbox{\tiny{UL}}},s_{\mbox{\tiny{UL}}})}\left(c(t_{\mbox{\tiny{UL}}},s_{\mbox{\tiny{UL}}})-\frac{1}{\lambda_i}\right),\label{eq:ul_energy}
\end{equation}
where the water-level can be calculated as
\begin{equation}
c(t_{\mbox{\tiny{UL}}},s_{\mbox{\tiny{UL}}})=\frac{2^{\frac{s_{\mbox{\tiny{UL}}}}{W_{\mbox{\tiny{UL}}}t_{\mbox{\tiny{UL}}}K(t_{\mbox{\tiny{UL}}},s_{\mbox{\tiny{UL}}})}}}{\left(\prod_{k=1}^{K(t_{\mbox{\tiny{UL}}},s_{\mbox{\tiny{UL}}})}\lambda_k\right)^{\frac{1}{K(t_{\mbox{\tiny{UL}}},s_{\mbox{\tiny{UL}}})}}}.\label{eq:constant_c}
\end{equation}

The number of active eigenmodes $K(t_{\mbox{\tiny{UL}}},s_{\mbox{\tiny{UL}}})\leq \operatorname{rank}\left(\mathbf{H}^H\mathbf{H}\right)$ can be calculated as follows: $K(t_{\mbox{\tiny{UL}}},s_{\mbox{\tiny{UL}}})=\operatorname{rank}\left(\mathbf{H}^H\mathbf{H}\right)$ if
\begin{equation}
\frac{2^{\frac{s_{\mbox{\tiny{UL}}}}{W_{\mbox{\tiny{UL}}}t_{\mbox{\tiny{UL}}}\operatorname{rank}\left(\mathbf{H}^H\mathbf{H}\right)}}}{\left(\prod_{k=1}^{\operatorname{rank}\left(\mathbf{H}^H\mathbf{H}\right)}\lambda_k\right)^{\frac{1}{\operatorname{rank}\left(\mathbf{H}^H\mathbf{H}\right)}}}>\frac{1}{\lambda_{\operatorname{rank}\left(\mathbf{H}^H\mathbf{H}\right)}};\label{eq:def_K_1}
\end{equation}
otherwise, $K(t_{\mbox{\tiny{UL}}},s_{\mbox{\tiny{UL}}})$ will be the value of $\widetilde{K}$ (with $1\leq
\widetilde{K}<\operatorname{rank}\left(\mathbf{H}^H\mathbf{H}\right)$) for which the following conditions hold:
\begin{equation}
\frac{2^{\frac{s_{\mbox{\tiny{UL}}}}{W_{\mbox{\tiny{UL}}}t_{\mbox{\tiny{UL}}}\widetilde{K}}}}{\left(\prod_{k=1}^{\widetilde{K}}\lambda_k\right)^{\frac{1}{\widetilde{K}}}}>\frac{1}{\lambda_{\widetilde{K}}}\hspace{0.5cm}
\operatorname{and}\hspace{0.5cm}\frac{2^{\frac{s_{\mbox{\tiny{UL}}}}{W_{\mbox{\tiny{UL}}}t_{\mbox{\tiny{UL}}}\widetilde{K}}}}{\left(\prod_{k=1}^{\widetilde{K}}\lambda_k\right)^{\frac{1}{\widetilde{K}}}}\leq\frac{1}{\lambda_{\widetilde{K}+1}}.\label{eq:def_K_2}
\end{equation}

In the SISO case (i.e., when both the
MT and the FAP have a single antenna), the minimum communication
energy resulting from problem (\ref{eq:energy_latency_power_ul}) and (\ref{eq:ul_energy})
is expressed as
\begin{equation}
e_{\mbox{\tiny{UL}}}(t_{\mbox{\tiny{UL}}},s_{\mbox{\tiny{UL}}})=k_{\mbox{\scriptsize{tx}},1}t_{\mbox{\tiny{UL}}}+k_{\mbox{\scriptsize{tx}},2}t_{\mbox{\tiny{UL}}}\frac{2^{\frac{s_{\mbox{\tiny{UL}}}}{W_{\mbox{\tiny{UL}}}t_{\mbox{\tiny{UL}}}}}-1}{\gamma_{\mbox{\tiny{UL}}}},
\end{equation}
where $\gamma_{\mbox{\tiny{UL}}}=|h_{\mbox{\tiny{UL}}}|^2$, being $h_{\mbox{\tiny{UL}}}$ the complex channel
gain between the MT and the FAP. The previous result can be proved easily taking into account that in the SISO case $K(t_{\mbox{\tiny{UL}}},s_{\mbox{\tiny{UL}}})=1$ and that the only channel eigenvalue is $\gamma_{\mbox{\tiny{UL}}}=|h_{\mbox{\tiny{UL}}}|^2$.

\subsubsection{Characterization}
As it will be shown in Section \ref{sec:joint_optim_radio_comput}, the minimum energy function $e_{\mbox{\tiny{UL}}}(t_{\mbox{\tiny{UL}}},s_{\mbox{\tiny{UL}}})$ will play a key role in the global resource allocation problem that includes communication and computation. Two important features of function $e_{\mbox{\tiny{UL}}}(t_{\mbox{\tiny{UL}}},s_{\mbox{\tiny{UL}}})$ are provided in the next two lemmas.

\begin{lemma}\label{lemma:eul_convex}
The minimum UL energy consumption function $e_{\mbox{\tiny{UL}}}(t_{\mbox{\tiny{UL}}},s_{\mbox{\tiny{UL}}})$ is jointly convex w.r.t. $t_{\mbox{\tiny{UL}}}$ and $s_{\mbox{\tiny{UL}}}$.
\end{lemma}

\begin{IEEEproof}
Problem (\ref{eq:energy_latency_power_ul}) is equivalent to the following convex optimization problem:
\begin{equation}
\begin{array}
[c]{rl}
\operatorname*{minimize}\limits_{\tau_{\mbox{\tiny{UL}}},\mathbf{Q}} & k_{\mbox{\scriptsize{tx}},1}\tau_{\mbox{\tiny{UL}}}+k_{\mbox{\scriptsize{tx}},2}\mbox{Tr}({\mathbf{Q}})\\
\operatorname*{subject\,\,to} &
\!\!\begin{array}
[t]{l} C1: s_{\mbox{\tiny{UL}}}\leq
W_{\mbox{\tiny{UL}}}\tau_{\mbox{\tiny{UL}}}\log_2\left|\mathbf{I}+\frac{\mathbf{H}\mathbf{Q}\mathbf{H}^H}{\tau_{\mbox{\tiny{UL}}}}\right|,
\\ C2: \tau_{\mbox{\tiny{UL}}}=t_{\mbox{\tiny{UL}}}, \\ C3:
\mathbf{Q}\succeq\mathbf{0},
\end{array}
\end{array}
\label{eq:energy_latency_tradeoff_ul}
\end{equation}
where $\tau_{\mbox{\tiny{UL}}}$ and the energy covariance matrix, defined as $\mathbf{Q}=\tau_{\mbox{\tiny{UL}}}\widetilde{\mathbf{Q}}$, are the optimization variables and $t_{\mbox{\tiny{UL}}}$ and $s_{\mbox{\tiny{UL}}}$ are parameters. Using a result from \cite{boyd:04}, the optimum value of the cost function in the above problem is convex w.r.t. the parameters $t_{\mbox{\tiny{UL}}}$ and $s_{\mbox{\tiny{UL}}}$. Finally, as the the solution of the above problem is the same one as the solution for problem (\ref{eq:energy_latency_power_ul}),\footnote{This is true since it can be verified that the optimum values of $\tau_{\mbox{\tiny{UL}}}$ and $\mathbf{Q}$ are $\tau_{\mbox{\tiny{UL}}}^{\star}=t_{\mbox{\tiny{UL}}}$ and $\mathbf{Q}^{\star}=t_{\mbox{\tiny{UL}}}\widetilde{\mathbf{Q}}^{\star}$.} it is concluded that $e_{\mbox{\tiny{UL}}}(t_{\mbox{\tiny{UL}}},s_{\mbox{\tiny{UL}}})$ is jointly convex w.r.t. $t_{\mbox{\tiny{UL}}}$ and $s_{\mbox{\tiny{UL}}}$.
\end{IEEEproof}

\begin{figure}[t]
    \begin{center}
        \includegraphics[width=11.5cm]{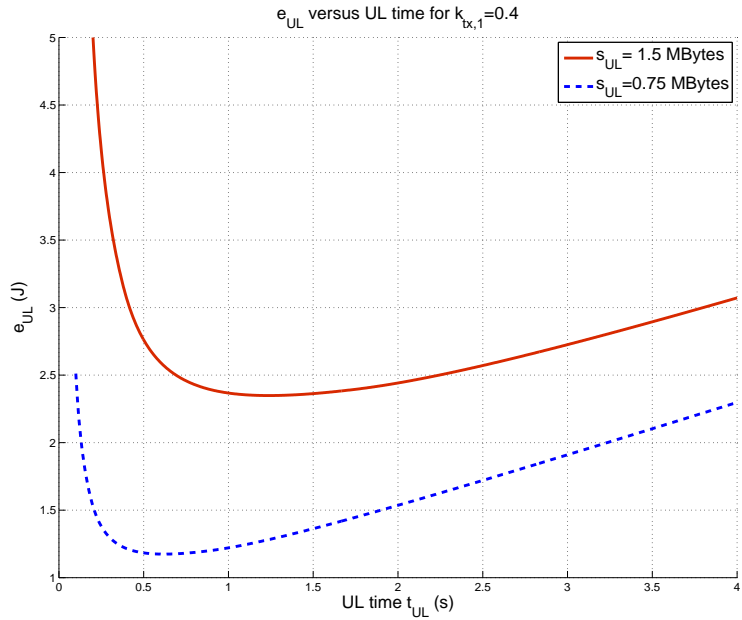}
    \end{center}
\vspace{-1.2cm}
    \caption{UL energy $e_{\mbox{\tiny{UL}}}(t_{\mbox{\tiny{UL}}},s_{\mbox{\tiny{UL}}})$ vs. transmission time $t_{\mbox{\tiny{UL}}}$ for $k_{\mbox{\scriptsize{tx}},1}=0.4$ W, $k_{\mbox{\scriptsize{tx}},2}=18$, and two different data block sizes: 1.5 and 0.75 MBytes. When $s_{\mbox{\tiny{UL}}}=1.5$ MBytes, the minimum is achieved at $t_{\mbox{\tiny{UL}}}=1.24$ s, whereas when $s_{\mbox{\tiny{UL}}}=0.75$ MBytes, the minimum is achieved at $t_{\mbox{\tiny{UL}}}=0.62$ s.}
    \label{fig:eUL_vs_time}
\end{figure}

\begin{figure}[t]
    \begin{center}
        \includegraphics[width=11.5cm]{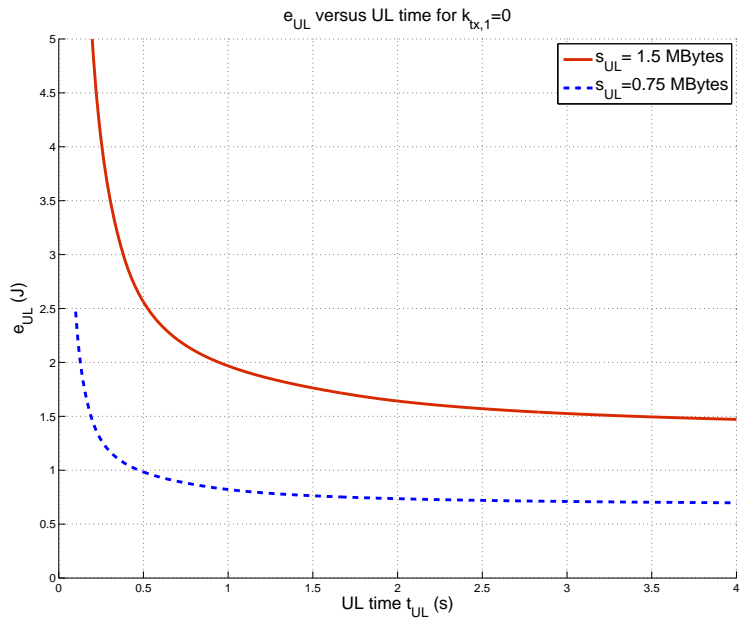}
    \end{center}
\vspace{-1.2cm}
    \caption{UL energy $e_{\mbox{\tiny{UL}}}(t_{\mbox{\tiny{UL}}},s_{\mbox{\tiny{UL}}})$ vs. transmission time $t_{\mbox{\tiny{UL}}}$ for $k_{\mbox{\scriptsize{tx}},1}=0$, $k_{\mbox{\scriptsize{tx}},2}=18$, and two different data block sizes: 1.5 and 0.75 MBytes.}
    \label{fig:eUL_vs_time_ktxcero}
\end{figure}

As a consequence from the joint convexity of function $e_{\mbox{\tiny{UL}}}(t_{\mbox{\tiny{UL}}},s_{\mbox{\tiny{UL}}})$ w.r.t. $t_{\mbox{\tiny{UL}}}$ and $s_{\mbox{\tiny{UL}}}$ (see Lemma \ref{lemma:eul_convex}), it can be concluded that for a given value of $s_{\mbox{\tiny{UL}}}$, the energy vs. time function will be also convex \cite{boyd:04}. Fig. \ref{fig:eUL_vs_time} and \ref{fig:eUL_vs_time_ktxcero} show the energy vs. UL time $t_{\mbox{\tiny{UL}}}$ considering two data block sizes to be transmitted through the UL ($s_{\mbox{\tiny{UL}}}=0.75$ and 1.75 MBytes), a concrete realization of a 4x4 MIMO channel with a bandwidth $W_{\mbox{\tiny{UL}}}=10$ MHz, and assuming $k_{\mbox{\scriptsize{tx}},2}=18$. In particular, Fig. \ref{fig:eUL_vs_time} corresponds to the case where $k_{\mbox{\scriptsize{tx}},1}=0.4$ W, i.e., the MT spends a non-negligible baseline power for having the transmission RF and BB circuitry switched on. On the other hand, in Fig. \ref{fig:eUL_vs_time_ktxcero} such constant is 0, which means that the only power that the MT spends comes from the radiated power. The main consequence from this is that in Fig. \ref{fig:eUL_vs_time} the curves present a minimum w.r.t. the UL time $t_{\mbox{\tiny{UL}}}$ (and, therefore, it does not make sense to spend more time than that corresponding to such a minimum), whereas in Fig. \ref{fig:eUL_vs_time_ktxcero} the spent energy decreases with the UL transmission time.\footnote{The concrete numerical results depend, of course, on the numerical values adopted for the parameters of the energy consumption models. Note also that some specific values have to be selected to produce the simulations. However, in Section IV we have deduced a set of generic characteristics of the curves relating the energy and the latency in the wireless transmission. More specifically, Section IV shows that, irrespectively of the numerical values of the model parameters, only two cases are possible, namely, the case in which the curve has a single minimum (as shown in Fig. \ref{fig:eUL_vs_time}) and the case in which the curve is monotonous decreasing (as shown in Fig. \ref{fig:eUL_vs_time_ktxcero}). Despite a different set of parameters will change the specific shape of the energy vs. UL transmission time curve shown in Fig. \ref{fig:eUL_vs_time} and \ref{fig:eUL_vs_time_ktxcero}, the resulting curve will belong, anyway, to one of the two described cases.\\It should be also emphasized that, although for the sake of analytical treatment we have considered the models (\ref{eq:ul_power_model}) and (\ref{eq:dl_power_model}) that simplify the general ones provided in \cite{jensen:12}, the values we have selected here for the parameters allow models to approximate quite well the extra energy consumption due to the offloading according to the experimental measurements provided in \cite{jensen:12} for a practical LTE handset.}

It should be also emphasized that, although the number of active eigenmodes $K(t_{\mbox{\tiny{UL}}},s_{\mbox{\tiny{UL}}})$ is a discrete function, $e_{\mbox{\tiny{UL}}}(t_{\mbox{\tiny{UL}}},s_{\mbox{\tiny{UL}}})$ and $c(t_{\mbox{\tiny{UL}}},s_{\mbox{\tiny{UL}}})$ are continuous w.r.t. $t_{\mbox{\tiny{UL}}}$ and $s_{\mbox{\tiny{UL}}}$. The reason behind this statement is that, at the instant in which a new eigenmode is activated due to an increase of the water-level, the power that is allocated to it is zero. Then, when the water-level keeps on increasing, the powers allocated to the activated eigenmodes also increase continuously.

\begin{lemma}\label{lemma:rate}
The UL energy normalized by the number of transmitted bits, i.e.,
$\frac{1}{s_{\mbox{\tiny{UL}}}}e_{\mbox{\tiny{UL}}}(t_{\mbox{\tiny{UL}}},s_{\mbox{\tiny{UL}}})$,
depends only on the UL rate
$r_{\mbox{\tiny{UL}}}=\frac{s_{\mbox{\tiny{UL}}}}{t_{\mbox{\tiny{UL}}}}$.
This allows to introduce the following notation:
$\overline{e}_{\mbox{\tiny{UL}}}(r_{\mbox{\tiny{UL}}})=\overline{e}_{\mbox{\tiny{UL}}}\left(\frac{s_{\mbox{\tiny{UL}}}}{t_{\mbox{\tiny{UL}}}}\right)=\frac{1}{s_{\mbox{\tiny{UL}}}}e_{\mbox{\tiny{UL}}}(t_{\mbox{\tiny{UL}}},s_{\mbox{\tiny{UL}}})=e_{\mbox{\tiny{UL}}}\left(\frac{1}{r_{\mbox{\tiny{UL}}}},1\right)$
or, equivalently,
$e_{\mbox{\tiny{UL}}}(t_{\mbox{\tiny{UL}}},s_{\mbox{\tiny{UL}}})=s_{\mbox{\tiny{UL}}}\overline{e}_{\mbox{\tiny{UL}}}(r_{\mbox{\tiny{UL}}})$.
In addition, function $\overline{e}_{\mbox{\tiny{UL}}}(r_{\mbox{\tiny{UL}}})$ is characterized by the fact that any local minimum will be also the global minimum of $\overline{e}_{\mbox{\tiny{UL}}}(r_{\mbox{\tiny{UL}}})$.
\end{lemma}

\begin{IEEEproof}
From expressions (\ref{eq:constant_c}), (\ref{eq:def_K_1}), and (\ref{eq:def_K_2}), it can be verified easily that functions $c(t_{\mbox{\tiny{UL}}},s_{\mbox{\tiny{UL}}})$ and $K(t_{\mbox{\tiny{UL}}},s_{\mbox{\tiny{UL}}})$ depend only on the rate $r_{\mbox{\tiny{UL}}}=\frac{s_{\mbox{\tiny{UL}}}}{t_{\mbox{\tiny{UL}}}}$. Based on this, the following notation will be used when considered appropriate:
$K(t_{\mbox{\tiny{UL}}},s_{\mbox{\tiny{UL}}})=K\left(\frac{s_{\mbox{\tiny{UL}}}}{t_{\mbox{\tiny{UL}}}}\right)=K(r_{\mbox{\tiny{UL}}})$
and
$c(t_{\mbox{\tiny{UL}}},s_{\mbox{\tiny{UL}}})=c\left(\frac{s_{\mbox{\tiny{UL}}}}{t_{\mbox{\tiny{UL}}}}\right)=c(r_{\mbox{\tiny{UL}}})$.
In addition, through an analysis of expressions (\ref{eq:def_K_1}) and (\ref{eq:def_K_2}), it can be concluded that $K(r_{\mbox{\tiny{UL}}})$ is monotonous increasing w.r.t.
$r_{\mbox{\tiny{UL}}}$.

From the previous observation and using (\ref{eq:ul_energy}),
it can be verified that the UL energy normalized by the number of
transmitted bits, i.e.,
$\frac{1}{s_{\mbox{\tiny{UL}}}}e_{\mbox{\tiny{UL}}}(t_{\mbox{\tiny{UL}}},s_{\mbox{\tiny{UL}}})$,
depends only on the UL rate
$r_{\mbox{\tiny{UL}}}=\frac{s_{\mbox{\tiny{UL}}}}{t_{\mbox{\tiny{UL}}}}$. Note that since function
$\overline{e}_{\mbox{\tiny{UL}}}(r_{\mbox{\tiny{UL}}})$ can be
expressed as
$e_{\mbox{\tiny{UL}}}\left(\frac{1}{r_{\mbox{\tiny{UL}}}},1\right)$,
being $e_{\mbox{\tiny{UL}}}$ a convex function, then any local
minimum of $\overline{e}_{\mbox{\tiny{UL}}}(r_{\mbox{\tiny{UL}}})$
will be unique and also the global minimum although $\overline{e}_{\mbox{\tiny{UL}}}(r_{\mbox{\tiny{UL}}})$
does not have to be necessarily convex w.r.t. $r_{\mbox{\tiny{UL}}}$.
\end{IEEEproof}

\begin{figure}[t]
    \begin{center}
        \includegraphics[width=11.5cm]{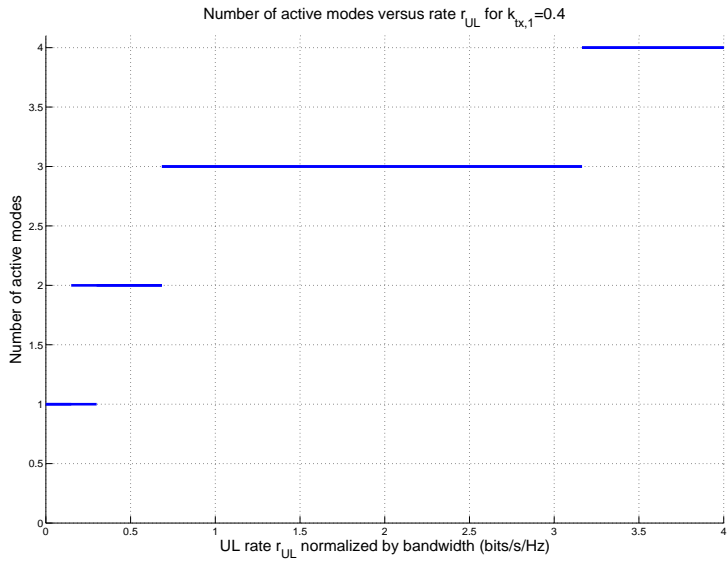}
    \end{center}
\vspace{-1.2cm}
    \caption{Number of active modes vs. $r_{\mbox{\tiny{UL}}}$ normalized by the UL bandwidth for $k_{\mbox{\scriptsize{tx}},1}=0.4$ W.}
    \label{fig:K_vs_r}
\end{figure}

As an illustrative example, Fig. \ref{fig:K_vs_r} shows the number of active modes $K(r_{\mbox{\tiny{UL}}})$ w.r.t. the UL rate $r_{\mbox{\tiny{UL}}}$ when $k_{\mbox{\scriptsize{tx}},1}=0.4$ W and taking the same parameters used to generate the previous figures. As can be observed, the number of modes increases with the rate until achieving the 4 spatial eigenmodes available for a 4x4 MIMO channel, as expected.

\begin{figure}[t]
    \begin{center}
        \includegraphics[width=11.5cm]{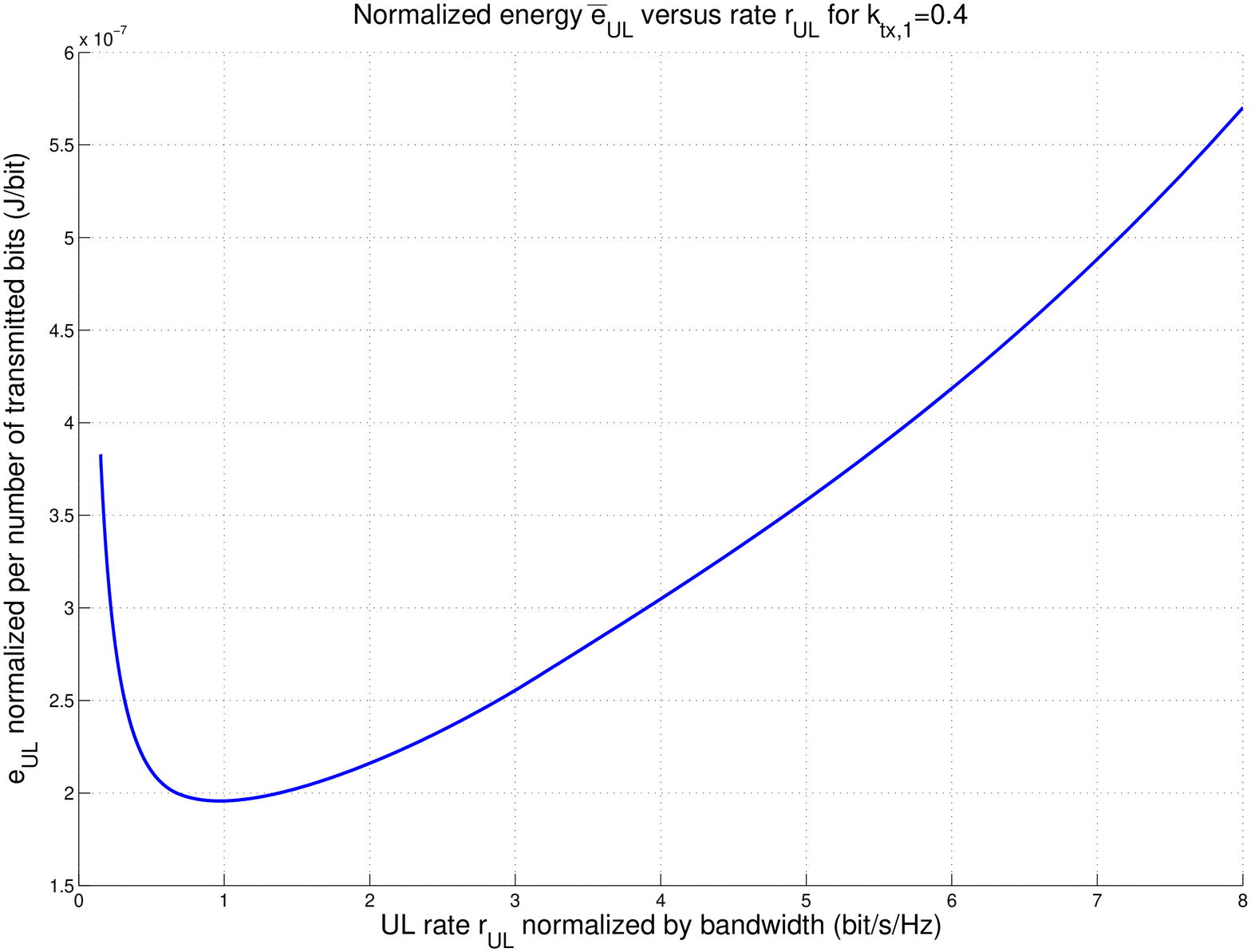}
    \end{center}
\vspace{-1.2cm}
    \caption{Normalized energy $\overline{e}_{\mbox{\tiny{UL}}}(r_{\mbox{\tiny{UL}}})$ vs. $r_{\mbox{\tiny{UL}}}$ normalized by the UL bandwidth for $k_{\mbox{\scriptsize{tx}},1}=0.4$ W.}
    \label{fig:N_eUL_vs_rate}
\end{figure}

\begin{figure}[t]
    \begin{center}
        \includegraphics[width=11.5cm]{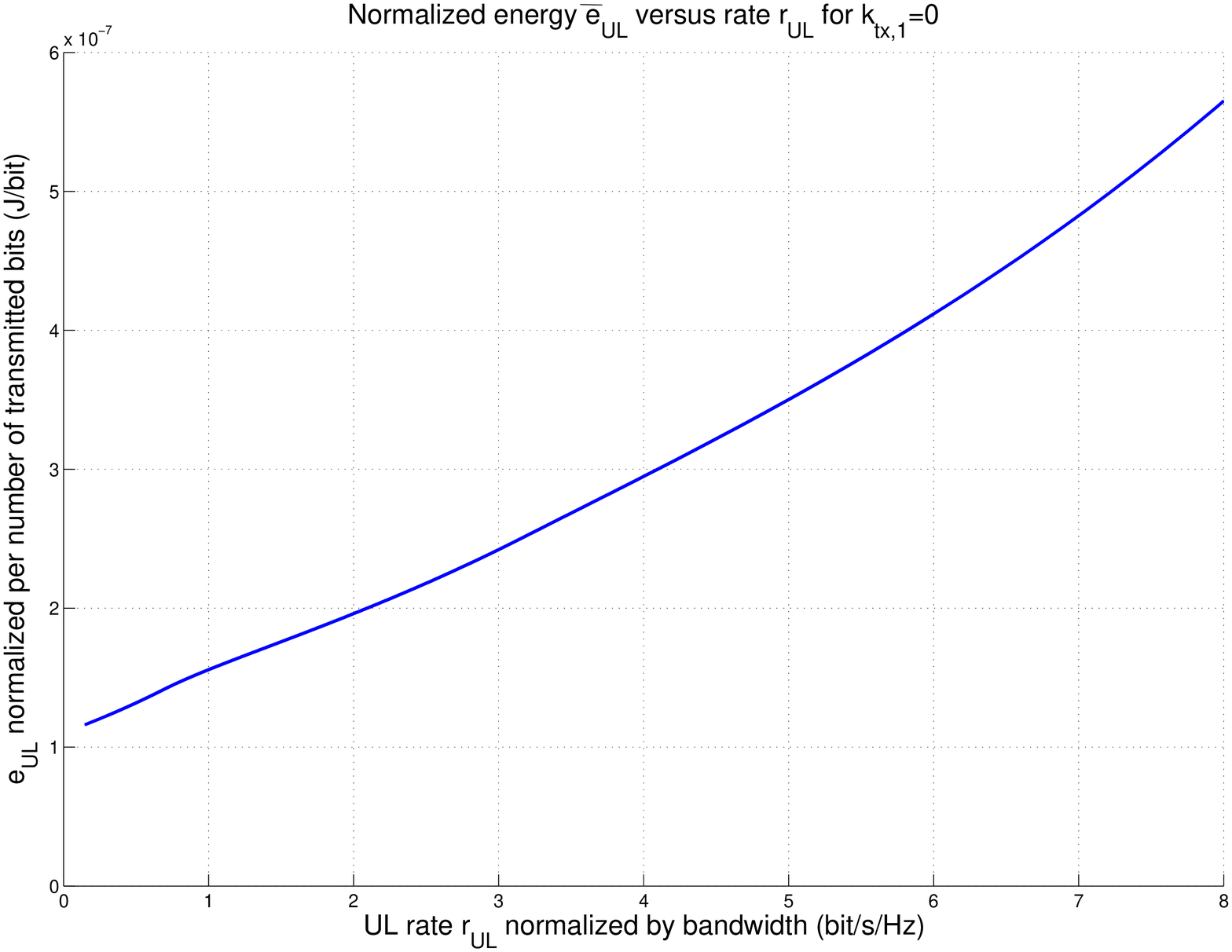}
    \end{center}
\vspace{-1.2cm}
    \caption{Normalized energy $\overline{e}_{\mbox{\tiny{UL}}}(r_{\mbox{\tiny{UL}}})$ vs. $r_{\mbox{\tiny{UL}}}$ normalized by the UL bandwidth for $k_{\mbox{\scriptsize{tx}},1}=0$.}
    \label{fig:N_eUL_vs_rate_ktxcero}
\end{figure}

Thanks to Lemma \ref{lemma:rate}, in the following, we will use
$\check{R}_{\mbox{\tiny{UL}}}$ to denote the UL rate that minimizes
$\overline{e}_{\mbox{\tiny{UL}}}(r_{\mbox{\tiny{UL}}})$. Note that
$\check{R}_{\mbox{\tiny{UL}}}\rightarrow\infty$ and
$\check{R}_{\mbox{\tiny{UL}}}=0$ means that the function
$\overline{e}_{\mbox{\tiny{UL}}}(r_{\mbox{\tiny{UL}}})$ is
monotonous decreasing and increasing, respectively. Fig.
\ref{fig:N_eUL_vs_rate} and \ref{fig:N_eUL_vs_rate_ktxcero} show the
normalized energy $\overline{e}_{\mbox{\tiny{UL}}}$ as a function of the UL transmission rate
$r_{\mbox{\tiny{UL}}}$ (normalized by the bandwidth) in the same
simulation conditions as the ones used to generate the previous
curves in this section (single realization of a 4x4 MIMO channel
with a bandwidth $W_{\mbox{\tiny{UL}}}=10$ MHz). Fig.
\ref{fig:N_eUL_vs_rate} corresponds to the case of
$k_{\mbox{\scriptsize{tx}},1}=0.4$ W, whereas in Fig.
\ref{fig:N_eUL_vs_rate_ktxcero}, $k_{\mbox{\scriptsize{tx}},1}=0$ is
considered. In the first case, the curve presents a minimum at
$\check{R}_{\mbox{\tiny{UL}}}=0.97$ b/s/Hz. Note that this
minimum could have been obtained without distinction by dividing the $s_{\mbox{\tiny{UL}}}$ by $t_{\mbox{\tiny{UL}}}$ values at any of the
minimums of the curves in Fig. \ref{fig:eUL_vs_time}. Note
also that in Fig. \ref{fig:N_eUL_vs_rate_ktxcero}, where we have assumed that $k_{\mbox{\scriptsize{tx}},1}=0$, the normalized
energy is monotonous increasing and, therefore,
$\check{R}_{\mbox{\tiny{UL}}}=0$.

\subsection{Trade-off between Latency and Energy in the DL Transmission}\label{subsec:lat_energ_tradeoff_dl}
In the case of the DL transmission, the relationship between the energy spent by the
MT when receiving ($e_{\mbox{\tiny{DL}}}$), the number of bits
transmitted through the DL ($s_{\mbox{\tiny{DL}}}$), and the time
spent in such DL transmission ($t_{\mbox{\tiny{DL}}}$) can be
formulated as follows based on (\ref{eq:dl_power_model}):
\begin{equation}
e_{\mbox{\tiny{DL}}}(t_{\mbox{\tiny{DL}}},s_{\mbox{\tiny{DL}}})=k_{\mbox{\scriptsize{rx}},1}t_{\mbox{\tiny{DL}}}+k_{\mbox{\scriptsize{rx}},2}s_{\mbox{\tiny{DL}}}.\label{eq:dl_total_energy_consump}
\end{equation}
The previous expression is linear and, thus, jointly
convex w.r.t. $t_{\mbox{\tiny{DL}}}$ and $s_{\mbox{\tiny{DL}}}$. As for the UL case, we may define the DL energy normalized by the number of received bits, which depends only on the DL rate $r_{\mbox{\tiny{DL}}}=\frac{s_{\mbox{\tiny{DL}}}}{t_{\mbox{\tiny{DL}}}}$: $\overline{e}_{\mbox{\tiny{DL}}}(r_{\mbox{\tiny{DL}}})=\overline{e}_{\mbox{\tiny{DL}}}\left(\frac{s_{\mbox{\tiny{DL}}}}{t_{\mbox{\tiny{DL}}}}\right)=\frac{1}{s_{\mbox{\tiny{DL}}}}e_{\mbox{\tiny{DL}}}(t_{\mbox{\tiny{DL}}},s_{\mbox{\tiny{DL}}})=\frac{k_{\mbox{\scriptsize{rx}},1}}{r_{\mbox{\tiny{DL}}}}+k_{\mbox{\scriptsize{rx}},2}$. As it can be seen, this is a decreasing function w.r.t. $r_{\mbox{\tiny{DL}}}$.

It is important to emphasize that given values of $t_{\mbox{\tiny{DL}}}$ and $s_{\mbox{\tiny{DL}}}$ will be feasible only if
the DL channel supports the rate $r_{\mbox{\tiny{DL}}}=\frac{s_{\mbox{\tiny{DL}}}}{t_{\mbox{\tiny{DL}}}}$. If
we consider that the FAP is endowed with $n_{\mbox{\tiny{FAP}}}$ antennas and that
only its radiated power is constrained by $P_{\mbox{\scriptsize{tx}},\mbox{\tiny{FAP}}}$, then the
following relationship has to be fulfilled:
\begin{equation}
r_{\mbox{\tiny{DL}}}=\frac{s_{\mbox{\tiny{DL}}}}{t_{\mbox{\tiny{DL}}}}\leq
W_{\mbox{\tiny{DL}}}\log_2\left|\mathbf{I}+\mathbf{H}_{\mbox{\tiny{DL}}}\mathbf{\widetilde{Q}}_{\mbox{\tiny{DL}}}\mathbf{H}_{\mbox{\tiny{DL}}}^H\right|\mbox{
for some }\mathbf{\widetilde{Q}}_{\mbox{\tiny{DL}}}\mbox{ with
}\mbox{Tr}({\mathbf{\widetilde{Q}}}_{\mbox{\tiny{DL}}})\leq P_{\mbox{\scriptsize{tx}},\mbox{\tiny{FAP}}},
\end{equation}
where $\mathbf{\widetilde{Q}}_{\mbox{\tiny{DL}}}$ represents the
transmit power covariance matrix at the transmitting serving FAP and
$\mathbf{H}_{\mbox{\tiny{DL}}}\in\mathds{C}^{n_{\mbox{\tiny{MT}}}\times
n_{\mbox{\tiny{FAP}}}}$ denotes the response of the MIMO channel in
DL.

The maximum supported DL rate can be calculated as the solution to
the following problem:
\begin{equation}
\begin{array}
[c]{cl}
\operatorname*{maximize}\limits_{\mathbf{\widetilde{Q}}_{\mbox{\tiny{DL}}}} & W_{\mbox{\tiny{DL}}}\log_2\left|\mathbf{I}+\mathbf{H}_{\mbox{\tiny{DL}}}\mathbf{\widetilde{Q}}_{\mbox{\tiny{DL}}}\mathbf{H}_{\mbox{\tiny{DL}}}^H\right|\\
\operatorname*{subject\,\,to} &
\!\!\begin{array}
[t]{l} C1: \operatorname{Tr}(\mathbf{\widetilde{Q}}_{\mbox{\tiny{DL}}})\leq
P_{\mbox{\scriptsize{tx}},\mbox{\tiny{FAP}}}, \\ C2: \mathbf{\widetilde{Q}}_{\mbox{\tiny{DL}}}\succeq\mathbf{0}.
\end{array}
\end{array}
\label{eq:max_rate_sl}
\end{equation}

The previous problem is convex (the objective function to be
maximized is concave) and the optimum solution consists in
transmitting through the eigenmodes of
$\mathbf{H}_{\mbox{\tiny{DL}}}^H\mathbf{H}_{\mbox{\tiny{DL}}}$ using
the well known water-filling over the corresponding eigenvalues
\cite{raleigh:98,palomar:03}. Accordingly, the optimum value of the
objective function, i.e., the maximum DL achievable rate, is
represented by $R_{\mbox{\tiny{DL}}}^{\max}$. Based on this, the
constraint to be fulfilled by the number of bits to be transmitted
in DL and the corresponding transmission time is
$r_{\mbox{\tiny{DL}}}=\frac{s_{\mbox{\tiny{DL}}}}{t_{\mbox{\tiny{DL}}}}\leq
R_{\mbox{\tiny{DL}}}^{\max}$.

\subsection{Main Conclusions}
In summary, the main results of this section are the following:

\begin{itemize}

\item To minimize the total energy consumed by the MT in the UL (or the energy normalized per transmitted bit), the UL transmission should be done through the channel eigenvectors, as expected. The number of active eigenmodes, upper bounded by the rank of $\mathbf{H}^H\mathbf{H}$, will depend only on the UL data rate and will be an increasing function of such rate. The total energy consumption per bit in the UL depends only on the UL data rate as well and presents a global minimum, even if the UL normalized energy is not a convex function. If the baseline energy consumption for having the transmitter chain switched on is negligible (i.e., $k_{tx,1}=0$), the energy consumption per bit in the UL is an increasing function of the UL data rate, which means that, if we want to minimize energy, we need to decrease the UL data rate as much as possible. However, if $k_{tx,1}$ is different from 0, then increasing the UL transmission time may not be the best solution after all. We will deal with this situation in the next section.

\item The total energy consumed by the MT in the DL per received bit depends only on the DL transmission rate and is a decreasing function of such rate. Therefore, to minimize the energy consumption by the MT in the DL, the optimal solution is that the FAP transmits with the highest possible DL rate $R_{\mbox{\tiny{DL}}}^{\max}$ that depends on the DL channel conditions and the maximum transmission power of the FAP.

\end{itemize}

\section{Joint Optimization of the Radio and Computational Resources with Partial
Offloading}\label{sec:joint_optim_radio_comput}

\subsection{Problem Formulation}
We address in this section the joint optimization of the usage of communication and computational resources in the offloading process, where a part of the processing will be done at the MT and a part will be offloaded to the FAP. When formulating the problem, several parameters, variables (see Table \ref{table:notation}), definitions, and aspects have to be taken into account, as detailed below.

\begin{table}[!t]
\renewcommand{\arraystretch}{1.1}
\caption{\textsc{Variables and Parameters Involved in the Offloading Optimization Problem}}
\vspace{-0.6cm} \label{table:notation}
\begin{center}
\begin{tabular}{|| l | l ||}
\hline \hline
$t_{\mbox{\tiny{UL}}}$ & Time duration of the UL transmission  \\ \hline
$t_{\mbox{\tiny{DL}}}$ & Time duration of the DL transmission  \\ \hline
$s_{\mbox{\tiny{UL}}}$ & Number of bits sent by the MT through the UL in $t_{\mbox{\tiny{UL}}}$ seconds  \\ \hline
$s_{\mbox{\tiny{DL}}}$ & Number of bits received by the MT through the DL in $t_{\mbox{\tiny{DL}}}$ seconds  \\ \hline
$r_{\mbox{\tiny{UL}}}$ & Transmission rate in the UL transmission ($r_{\mbox{\tiny{UL}}}=\frac{s_{\mbox{\tiny{UL}}}}{t_{\mbox{\tiny{UL}}}}$)  \\ \hline
$r_{\mbox{\tiny{DL}}}$ & Transmission rate in the DL transmission ($r_{\mbox{\tiny{DL}}}=\frac{s_{\mbox{\tiny{DL}}}}{t_{\mbox{\tiny{DL}}}}$)  \\ \hline
$S_{\mbox{\scriptsize{app}}}$ & Application load measured as the number of bits to be processed  \\ \hline
$S_{P_0}$ & Processing load at the MT, measured as the number of bits to be processed locally at the MT  \\ \hline
$S_{P_1}$ & Processing load at the FAP, measured as the number of bits to be processed remotely at the FAP  \\ \hline
$\beta_{\mbox{\tiny{UL}}}$ & It accounts for the overhead due to the UL communication, i.e., $s_{\mbox{\tiny{UL}}}=\beta_{\mbox{\tiny{UL}}}S_{P_1}$  \\ \hline
$\beta_{\mbox{\tiny{DL}}}$ & It accounts jointly for the overhead due to the DL communication and the ratio
between output\\ & and input bits associated to the execution of the remote process at the FAP, i.e., $s_{\mbox{\tiny{DL}}}=\beta_{\mbox{\tiny{DL}}}S_{P_1}$  \\ \hline
$\tau_{P_0}$ & Required computation time per bit processed locally at the MT  \\ \hline
$\tau_{P_1}$ & Required computation time per bit processed remotely at the FAP  \\ \hline
$\varepsilon_{P_0}$ & Energy consumed per bit processed locally at the MT  \\ \hline
$\varepsilon_{P_1}$ & Energy consumed per bit processed remotely at the FAP  \\ \hline
$L_{\max}$ & Maximum admissible latency in the execution of the application  \\ \hline
$P_{\mbox{\scriptsize{tx}},\mbox{\tiny{MT}}}$ & Maximum radiated power of the MT  \\ \hline
$P_{\mbox{\scriptsize{tx}},\mbox{\tiny{FAP}}}$ & Maximum radiated power of the FAP  \\ \hline
$R_{\mbox{\tiny{UL}}}^{\max}$ & Maximum data rate supported in the UL transmission \\ \hline
$R_{\mbox{\tiny{DL}}}^{\max}$ & Maximum data rate supported in the DL transmission \\ \hline
$k_{\mbox{\scriptsize{tx}},1},k_{\mbox{\scriptsize{tx}},2},k_{\mbox{\scriptsize{rx}},1},k_{\mbox{\scriptsize{rx}},2}$ & Model dependent constants for the computation of the energy consumption in both UL and DL \\ \hline
$e_{\mbox{\tiny{UL}}}(t_{\mbox{\tiny{UL}}},s_{\mbox{\tiny{UL}}})$ & Energy spent by the MT when
transmitting through the UL (it is a function of $t_{\mbox{\tiny{UL}}}$ and $s_{\mbox{\tiny{UL}}}$) \\ \hline
$e_{\mbox{\tiny{DL}}}(t_{\mbox{\tiny{DL}}},s_{\mbox{\tiny{DL}}})$ & Energy spent by the MT when
receiving through the DL (it is a function of $t_{\mbox{\tiny{DL}}}$ and $s_{\mbox{\tiny{DL}}}$) \\ \hline
$\overline{e}_{\mbox{\tiny{UL}}}(r_{\mbox{\tiny{UL}}})$ & Normalized consumed energy per transmitted bit through the UL (it is a function of $r_{\mbox{\tiny{UL}}}$) \\ \hline
$\overline{e}_{\mbox{\tiny{DL}}}(r_{\mbox{\tiny{DL}}})$ & Normalized consumed energy per received bit through the DL (it is a function of $r_{\mbox{\tiny{DL}}}$) \\ \hline
$\check{R}_{\mbox{\tiny{UL}}}$ & Value of $r_{\mbox{\tiny{UL}}}$ for which the normalized energy in the UL $\overline{e}_{\mbox{\tiny{UL}}}(r_{\mbox{\tiny{UL}}})$ is minimized \\ \hline
$c(r_{\mbox{\tiny{UL}}})$ & Water-level for the computation of the power assigned to each active eigenmode in the UL\\ & (it is a function of $r_{\mbox{\tiny{UL}}}$) \\ \hline
$K(r_{\mbox{\tiny{UL}}})$ & Number of active eigenmodes in the UL (it is a function of $r_{\mbox{\tiny{UL}}}$) \\
\hline \hline
\end{tabular}
\end{center}
\vspace{-0.8cm}
\end{table}

%[c]{cl}
%\operatorname*{minimize}\limits_{S_{P_0},S_{P_1},t_{\mbox{\tiny{UL}}},t_{\mbox{\tiny{DL}}}} & e_{\mbox{\tiny{UL}}}(t_{\mbox{\tiny{UL}}},\beta_{\mbox{\tiny{UL}}}S_{P_1})+\varepsilon_{P_0}S_{P_0}+e_{\mbox{\tiny{DL}}}(t_{\mbox{\tiny{DL}}},\beta_{\mbox{\tiny{DL}}}S_{P_1}) \\
%\operatorname*{subject\,\,to} &
%\!\!\begin{array}
%[t]{l} C1: S_{P_0}+S_{P_1}=S_{\mbox{\scriptsize{app}}},\\C2:
%\max\left\{\tau_{P_0}S_{P_0},t_{\mbox{\tiny{UL}}}+\tau_{P_1}S_{P_1}+t_{\mbox{\tiny{DL}}}\right\}\leq
%L_{\max},\\C3: e_{\mbox{\tiny{UL}}}(t_{\mbox{\tiny{UL}}},\beta_{\mbox{\tiny{UL}}}S_{P_1})-k_{\mbox{\scriptsize{tx}},1}t_{\mbox{\tiny{UL}}}\leq
%k_{\mbox{\scriptsize{tx}},2}t_{\mbox{\tiny{UL}}}P_{\mbox{\scriptsize{tx}},\mbox{\tiny{MT}}},\\C4: \beta_{\mbox{\tiny{DL}}}S_{P_1}\leq
%t_{\mbox{\tiny{DL}}}R_{\mbox{\tiny{DL}}}^{\max}.
%\end{array}
%\end{array}
%\label{eq:global_optim_problem}
%\end{equation}

The ultimate goal is to minimize the total energy spent by the MT. Such energy includes the energy spent in the UL transmission and DL reception, as well as the energy spent in the local processing (see the objective function in the offloading optimization problem (\ref{eq:global_optim_problem})). On the other hand, the execution of the application has to finish within a time frame not longer than $L_{\max}$ associated to a given QoS to be perceived by the user.

Let us consider that the application to be executed has to process $S_{\mbox{\scriptsize{app}}}$ bits. We assume that these bits can be divided into two groups of any size, so that $S_{P_0}$ bits will be processed locally at the MT and $S_{P_1}$ bits will be processed remotely at the FAP. Although in a practical case, only some sizes may be accepted in the data partitioning, we take this approach in order to understand the fundamental tradeoffs in the offloading process. It is considered that both computation processes at the MT and the FAP can be performed in parallel and, for the sake of simplicity in the notation, we will assume that such division does not imply any overhead, i.e., $S_{\mbox{\scriptsize{app}}}=S_{P_0}+S_{P_1}$ (formulated as constraint C1 in (\ref{eq:global_optim_problem})).

Concerning the computational capabilities of the MT and the FAP, we denote the time that the MT and the FAP need to process a single bit by $\tau_{P_0}$ and $\tau_{P_1}$, respectively. Note that these parameters account jointly for the CPU rate (in cycles/second) and the complexity (in cycles/bit) associated to the application \cite{miettinen:10}. The time required for the execution of the application, i.e., the latency, will be given as the maximum value of the time required by the MT to perform the assigned local computation and the time required for the offloading. Such offloading time includes the transmission of the offloaded bits through the UL, the remote execution at the FAP, and the reception through the DL. This latency must be less than or equal to $L_{\max}$ (see constraint C2 in (\ref{eq:global_optim_problem})).

We assume that the number of bits to be transmitted through the UL is proportional to $S_{P_1}$ ($s_{\mbox{\tiny{UL}}}=\beta_{\mbox{\tiny{UL}}}S_{P_1}$), where the constant $\beta_{\mbox{\tiny{UL}}}>1$ accounts for the overhead due to the UL communication. Similarly, we assume that the number of bits to be transmitted through the DL is proportional to $S_{P_1}$ ($s_{\mbox{\tiny{DL}}}=\beta_{\mbox{\tiny{DL}}}S_{P_1}$), where the constant $\beta_{\mbox{\tiny{DL}}}$ accounts jointly for the overhead due to the DL communication and the ratio between output and input bits associated to the execution of the remote process at the FAP.

Both the MT and the FAP have a limitation in terms of maximum radiated power, represented by $P_{\mbox{\scriptsize{tx}},\mbox{\tiny{MT}}}$ and $P_{\mbox{\scriptsize{tx}},\mbox{\tiny{FAP}}}$, respectively, and introduced through constraints C3 and C4 in (\ref{eq:global_optim_problem}). Note that in addition to the communication itself, the MT also spends some energy in processing the bits related to the part of the application that is not offloaded. Such energy is modeled as $\varepsilon_{P_0}S_{P_0}$, where $\varepsilon_{P_0}$ represents the energy spent for each bit that has to be processed locally. This parameter accounts jointly for the energy/cycle of the MT processor and the cycles/bit associated to the application \cite{miettinen:10}.

Based on all the previous points, the resource allocation problem can be written as\footnote{Note that in case that we would like to incorporate economic related aspects, such as, for example, the potential payment for the use of the transmission link and/or the use of the FAP for remote computation, the formulation of the resource allocation problem should be adapted by modifying the cost function and/or adding new constraints accordingly. In case that we would like to include the energy consumption of the FAP, an additional term in the cost function and/or an additional constraint should be added accounting for this.}
\begin{equation}
\begin{array}
[c]{cl}
\operatorname*{minimize}\limits_{S_{P_0},S_{P_1},t_{\mbox{\tiny{UL}}},t_{\mbox{\tiny{DL}}}} & e_{\mbox{\tiny{UL}}}(t_{\mbox{\tiny{UL}}},\beta_{\mbox{\tiny{UL}}}S_{P_1})+\varepsilon_{P_0}S_{P_0}+e_{\mbox{\tiny{DL}}}(t_{\mbox{\tiny{DL}}},\beta_{\mbox{\tiny{DL}}}S_{P_1}) \\
\operatorname*{subject\,\,to} &
\!\!\begin{array}
[t]{l} C1: S_{P_0}+S_{P_1}=S_{\mbox{\scriptsize{app}}},\\C2:
\max\left\{\tau_{P_0}S_{P_0},t_{\mbox{\tiny{UL}}}+\tau_{P_1}S_{P_1}+t_{\mbox{\tiny{DL}}}\right\}\leq
L_{\max},\\C3: e_{\mbox{\tiny{UL}}}(t_{\mbox{\tiny{UL}}},\beta_{\mbox{\tiny{UL}}}S_{P_1})-k_{\mbox{\scriptsize{tx}},1}t_{\mbox{\tiny{UL}}}\leq
k_{\mbox{\scriptsize{tx}},2}t_{\mbox{\tiny{UL}}}P_{\mbox{\scriptsize{tx}},\mbox{\tiny{MT}}},\\C4: \beta_{\mbox{\tiny{DL}}}S_{P_1}\leq
t_{\mbox{\tiny{DL}}}R_{\mbox{\tiny{DL}}}^{\max}.
\end{array}
\end{array}
\label{eq:global_optim_problem}
\end{equation}

The previous problem is convex as the objective function is the sum of three functions that are either jointly convex or linear w.r.t. to the optimization variables, the inequality constraints are convex, and the equality constraints are linear \cite{boyd:04}.

\subsection{Simplification of the Global Resource Allocation Problem}
The objective now is to simplify the previous problem by
reformulating some of the previous constraints and by finding
partial solutions. The simplification is based on the following
facts:

\begin{itemize}

\item Thanks to C1 (the constraint that indicates that the total number of bits is distributed between local and remote processing), it is possible to express $S_{P_0}$ in terms of
$S_{P_1}$ as $S_{P_0}=S_{\mbox{\scriptsize{app}}}-S_{P_1}$. This will allow to eliminate
$S_{P_0}$ from the set of optimization variables.

\item Constraint C3 is the analytic formulation of the maximum radiated power at the MT associated to the UL transmission and, according to (\ref{eq:ul_energy}), this can be written as
\begin{equation}
\sum_{i=1}^{K(r_{\mbox{\tiny{UL}}})}\left(c(r_{\mbox{\tiny{UL}}})-\frac{1}{\lambda_i}\right)\leq
P_{\mbox{\scriptsize{tx}},\mbox{\tiny{MT}}}.\label{eq:reformulated_maximum_ul_rate}
\end{equation}
Taking into account that both $K(r_{\mbox{\tiny{UL}}})$ and $c(r_{\mbox{\tiny{UL}}})$ are
monotonous increasing functions, the previous constraint can be
written equivalently as
\begin{equation}
r_{\mbox{\tiny{UL}}}=\frac{s_{\mbox{\tiny{UL}}}}{t_{\mbox{\tiny{UL}}}}\leq R_{\mbox{\tiny{UL}}}^{\max},\label{eq:reformulated_maximum_ul_rate2}
\end{equation}
where $R_{\mbox{\tiny{UL}}}^{\max}$ is the UL rate for which
(\ref{eq:reformulated_maximum_ul_rate}) is fulfilled with equality.

\item As far as C4 is concerned (i.e., the constraint related to the maximum achievable rate in the DL transmission), in the optimum solution such constraint has
to be fulfilled with equality since, otherwise, we could always
decrease $t_{\mbox{\tiny{DL}}}$ until C4 is fulfilled with equality while, at the
same time, the objective function is reduced and constraint C2 may
become looser. Consequently, in the optimum solution we have:
\begin{equation}
t_{\mbox{\tiny{DL}}}=\frac{\beta_{\mbox{\tiny{DL}}}S_{P_1}}{R_{\mbox{\tiny{DL}}}^{\max}}.
\end{equation}
The previous equality will allow to eliminate variable $t_{\mbox{\tiny{DL}}}$ from
the set of optimization variables in the new simplified optimization
problem. Remember that the value of $R_{\mbox{\tiny{DL}}}^{\max}$ is directly related to the maximum power radiated by the FAP $P_{\mbox{\scriptsize{tx}},\mbox{\tiny{FAP}}}$, as explained in Subsection \ref{subsec:lat_energ_tradeoff_dl}.

\item Constraint C2 related to the available time budget (and formulated in terms of a maximum allowed
latency) can be rewritten as a set of two constraints
detailed as follows (where we have used the previous result
concerning the equality in C4 for the optimum solution):
\begin{equation}
C2a: \tau_{P_0}S_{P_0}=\tau_{P_0}(S_{\mbox{\scriptsize{app}}}-S_{P_1})\leq
L_{\max}\;\;\;\Rightarrow\;\;\;S_{P_1}\geq
S_{\mbox{\scriptsize{app}}}-\frac{L_{\max}}{\tau_{P_0}},\label{eq:const_c2a}
\end{equation}
\begin{eqnarray}
& C2b:
t_{\mbox{\tiny{UL}}}+\tau_{P_1}S_{P_1}+t_{\mbox{\tiny{DL}}}=\frac{\beta_{\mbox{\tiny{UL}}}S_{P_1}}{r_{\mbox{\tiny{UL}}}}+\tau_{P_1}S_{P_1}+\frac{\beta_{\mbox{\tiny{DL}}}S_{P_1}}{R_{\mbox{\tiny{DL}}}^{\max}}\leq
L_{\max}\\ &
\;\;\;\;\;\;\;\;\;\;\;\;\;\;\;\;\;\;\;\;\;\;\;\;\;\;\;\;\;\;\Rightarrow\;\;\;r_{\mbox{\tiny{UL}}}\geq
\frac{\beta_{\mbox{\tiny{UL}}}}{\frac{L_{\max}}{S_{P_1}}-\tau_{P_1}-\frac{\beta_{\mbox{\tiny{DL}}}}{R_{\mbox{\tiny{DL}}}^{\max}}}=\frac{\beta_{\mbox{\tiny{UL}}}S_{P_1}}{L_{\max}-\tau_{P_1}S_{P_1}-\frac{\beta_{\mbox{\tiny{DL}}}}{R_{\mbox{\tiny{DL}}}^{\max}}S_{P_1}}=r_{\mbox{\tiny{UL}}}^{\min}(S_{P_1}).\label{eq:def_rul_min}
\end{eqnarray}
Note that, in order to be able to find a feasible value of $r_{\mbox{\tiny{UL}}}$, we require that $r_{\mbox{\tiny{UL}}}^{\min}(S_{P_1})\leq
R_{\mbox{\tiny{UL}}}^{\max}$. Using (\ref{eq:def_rul_min}), this implies that the following condition on $S_{P_1}$ has to be fulfilled:
\begin{equation}
S_{P_1}
\leq\frac{L_{\max}}{\frac{\beta_{\mbox{\tiny{UL}}}}{R_{\mbox{\tiny{UL}}}^{\max}}+\tau_{P_1}+\frac{\beta_{\mbox{\tiny{DL}}}}{R_{\mbox{\tiny{DL}}}^{\max}}}.\label{eq:max_value_sp1}
\end{equation}

\end{itemize}

Using the previous results in (\ref{eq:const_c2a}) and (\ref{eq:max_value_sp1}), we define the minimum and maximum values of variable $S_{P_1}$ as follows:
\begin{eqnarray}
S_{P_1} & \geq & S_{P_1}^{\min},\;\;\;\;
S_{P_1}^{\min}=\max\left\{0,S_{\mbox{\scriptsize{app}}}-\frac{L_{\max}}{\tau_{P_0}}\right\},\label{eq:sp1_min}\\S_{P_1}
& \leq & S_{P_1}^{\max},\;\;\;\;
S_{P_1}^{\max}=\min\left\{S_{\mbox{\scriptsize{app}}},\frac{L_{\max}}{\frac{\beta_{\mbox{\tiny{UL}}}}{R_{\mbox{\tiny{UL}}}^{\max}}+\tau_{P_1}+\frac{\beta_{\mbox{\tiny{DL}}}}{R_{\mbox{\tiny{DL}}}^{\max}}}\right\}.\label{eq:sp1_max}
\end{eqnarray}

Taking all this into account, the optimization problem can be
expressed in a simplified way as follows (where we have reduced the
set of optimization variables to just two variables: $S_{P_1}$ and
$r_{\mbox{\tiny{UL}}}$):

\begin{equation}
\begin{array}
[c]{cl}
\operatorname*{minimize}\limits_{S_{P_1},r_{\mbox{\tiny{UL}}}} & S_{P_1}\beta_{\mbox{\tiny{UL}}}\overline{e}_{\mbox{\tiny{UL}}}(r_{\mbox{\tiny{UL}}})+\left(k_{\mbox{\scriptsize{rx}},1}\frac{\beta_{\mbox{\tiny{DL}}}}{R_{\mbox{\tiny{DL}}}^{\max}}+k_{\mbox{\scriptsize{rx}},2}\beta_{\mbox{\tiny{DL}}}-\varepsilon_{P_0}\right)S_{P_1}+\varepsilon_{P_0}S_{\mbox{\scriptsize{app}}} \\
\operatorname*{subject\,\,to} &
\!\!\begin{array}
[t]{l} S_{P_1}^{\min} \leq S_{P_1} \leq S_{P_1}^{\max},
\\
r_{\mbox{\tiny{UL}}}^{\min}(S_{P_1}) \leq  r_{\mbox{\tiny{UL}}}   \leq   R_{\mbox{\tiny{UL}}}^{\max}.
\end{array}
\end{array}
\label{eq:global_optim_problem_simple}
\end{equation}
The objective function in the previous problem is the same as the one in (\ref{eq:global_optim_problem}) after expressing all the optimization variables in terms of $S_{P_1}$ and $r_{\mbox{\tiny{UL}}}$, and formulating the UL energy consumption as $S_{P_1}\beta_{\mbox{\tiny{UL}}}\overline{e}_{\mbox{\tiny{UL}}}(r_{\mbox{\tiny{UL}}})$ and the DL energy consumption as $\left(k_{\mbox{\scriptsize{rx}},1}\frac{\beta_{\mbox{\tiny{DL}}}}{R_{\mbox{\tiny{DL}}}^{\max}}+k_{\mbox{\scriptsize{rx}},2}\beta_{\mbox{\tiny{DL}}}\right)S_{P_1}$ based on (\ref{eq:dl_total_energy_consump}).

The previous problem will be feasible if, and only if,
$S_{P_1}^{\min} \leq S_{P_1}^{\max}$. Otherwise, if the problem is infeasible, the only solution
is to increase the value of the maximum allowed latency $L_{\max}$.
In fact, using the previous expressions (\ref{eq:sp1_min}) and
(\ref{eq:sp1_max}), finding the value of $L_{\max}$ for which the
problem becomes feasible will be an easy task since $S_{P_1}^{\min}$
and $S_{P_1}^{\max}$ depend linearly on $L_{\max}$.

\subsection{Problem Solution}

In order to solve the previous problem, we will first of all
optimize variable $r_{\mbox{\tiny{UL}}}$ (i.e., the UL data rate) assuming a fixed
value of the number of bits $S_{P_1}$, obtaining as a result
$r_{\mbox{\tiny{UL}}}^{\star}(S_{P_1})$. Then, the remaining task will be the
optimization of variable $S_{P_1}$, which can be expressed as a
one-dimensional optimization problem and will be solved numerically
by means of an iterative procedure.

$r_{\mbox{\tiny{UL}}}^{\star}(S_{P_1})$ is found as the solution of the following
problem (note that, for a fixed value of the UL number of bits
$S_{P_1}$, the rate that minimizes the energy cost function in (\ref{eq:global_optim_problem_simple}) is equal to the rate
that minimizes the function $\overline{e}_{\mbox{\tiny{UL}}}(r_{\mbox{\tiny{UL}}})$ subject to the
constraints detailed below):
\begin{equation}
\begin{array}
[c]{rrl} r_{\mbox{\tiny{UL}}}^{\star}(S_{P_1}) = & \arg
\operatorname*{minimize}\limits_{r_{\mbox{\tiny{UL}}}} & \overline{e}_{\mbox{\tiny{UL}}}(r_{\mbox{\tiny{UL}}})
\\ &\operatorname*{subject\,\,to} &
\!\!\begin{array}
[t]{l} r_{\mbox{\tiny{UL}}}^{\min}(S_{P_1}) \leq  r_{\mbox{\tiny{UL}}} \leq R_{\mbox{\tiny{UL}}}^{\max}.
\end{array}
\end{array}
\label{eq:problem_optim_ul_rate}
\end{equation}

The solution to this problem is summarized as follows (recall that
$\overline{e}_{\mbox{\tiny{UL}}}(r_{\mbox{\tiny{UL}}})$ is a continuous function with a unique
minimum denoted by $\check{R}_{\mbox{\tiny{UL}}}$ so that for
$r_{\mbox{\tiny{UL}}}<\check{R}_{\mbox{\tiny{UL}}}$ the function is decreasing and for
$r_{\mbox{\tiny{UL}}}>\check{R}_{\mbox{\tiny{UL}}}$ the function is increasing):
\begin{equation}
r_{\mbox{\tiny{UL}}}^{\star}(S_{P_1})=\left\{\begin{array}[c]{ll} r_{\mbox{\tiny{UL}}}^{\min}(S_{P_1}), & \check{R}_{\mbox{\tiny{UL}}}<r_{\mbox{\tiny{UL}}}^{\min}(S_{P_1}),\\
\check{R}_{\mbox{\tiny{UL}}}, & r_{\mbox{\tiny{UL}}}^{\min}(S_{P_1})\leq \check{R}_{\mbox{\tiny{UL}}}\leq
R_{\mbox{\tiny{UL}}}^{\max}, \\ R_{\mbox{\tiny{UL}}}^{\max},& \check{R}_{\mbox{\tiny{UL}}}
> R_{\mbox{\tiny{UL}}}^{\max}.
\end{array}\right.\label{eq:r_ul_star}
\end{equation}
Note that, in the previous expression, each of the three lines corresponds to the case in which $\check{R}_{\mbox{\tiny{UL}}}$ lies on the left, within, or on the right of the search interval $\left[r_{\mbox{\tiny{UL}}}^{\min}(S_{P_1}),R_{\mbox{\tiny{UL}}}^{\max}\right]$.

Taking the previous result (\ref{eq:r_ul_star}) into account, the optimization problem (\ref{eq:global_optim_problem_simple}) can be rewritten as a simplified one-dimensional problem in terms of variable $S_{P_1}$:
\begin{equation}
\begin{array}
[c]{cl}
\operatorname*{minimize}\limits_{S_{P_1}} & f_o(S_{P_1}) \\
\operatorname*{subject\,\,to} &
\!\!\begin{array}
[t]{l} S_{P_1}^{\min} \leq S_{P_1} \leq S_{P_1}^{\max},
\end{array}
\end{array}
\label{eq:global_optim_problem_simple2}
\end{equation}
where the objective function $f_o(S_{P_1})$ is defined as
\begin{equation}
f_o(S_{P_1})=S_{P_1}\beta_{\mbox{\tiny{UL}}}\overline{e}_{\mbox{\tiny{UL}}}(r_{\mbox{\tiny{UL}}}^{\star}(S_{P_1}))+\left(k_{\mbox{\scriptsize{rx}},1}\frac{\beta_{\mbox{\tiny{DL}}}}{R_{\mbox{\tiny{DL}}}^{\max}}+k_{\mbox{\scriptsize{rx}},2}\beta_{\mbox{\tiny{DL}}}-\varepsilon_{P_0}\right)S_{P_1}+\varepsilon_{P_0}S_{\mbox{\scriptsize{app}}}.\label{eq:cost_function_fo_sp1}
\end{equation}

Note that the objective function in the previous problem (which is a
function of the single variable $S_{P_1}$) is numerically the same
as the one that we would have obtained by optimizing problem
(\ref{eq:global_optim_problem}) w.r.t. all the optimization
variables except $S_{P_1}$. The main consequence from this
observation is that, since (\ref{eq:global_optim_problem}) is a
convex optimization problem, the cost function $f_o(S_{P_1})$
(\ref{eq:cost_function_fo_sp1}) is a convex function w.r.t.
$S_{P_1}$. That allows to apply very simple numerical methods to solve problem (\ref{eq:global_optim_problem_simple2}) and calculate $S_{P_1}^{\star}$, that is, the optimum value of $S_{P_1}$ according to problem (\ref{eq:global_optim_problem_simple2}). Some illustrative examples of numerical methods are the gradient-based algorithms or the nested intervals technique \cite{quarteroni:07,suli:03}.

Table \ref{alg} presents the detailed steps of a numerical method that converges always with exponential speed \cite{quarteroni:07} and finds the optimum value $S_{P_1}^{\star}$ with a resolution better than a given percentage (represented by $\epsilon$) of the search interval length $S_{P_1}^{\max}-S_{P_1}^{\min}$. Note that steps 3-5 identify if the problem is
infeasible, whereas steps 6-10 and 11-15 check whether
$S_{P_1}^{\min}$ and $S_{P_1}^{\max}$ are the optimum solutions to
the problem, respectively.

The conditions under which $S_{P_1}^{\min}$ and $S_{P_1}^{\max}$ are the optimum solutions are derived by taking into account that, as it has been mentioned before, function $f_o(S_{P_1})$ is convex. Thanks to this observation, $S_{P_1}^{\star}=S_{P_1}^{\min}$ will be optimum if $\frac{df_o(S_{P_1}^{\min})}{dS_{P_1}}\geq 0$. On the other hand, the optimum solution will be $S_{P_1}^{\star}=S_{P_1}^{\max}$ if $\frac{df_o(S_{P_1}^{\max})}{dS_{P_1}}\leq 0$. Note that, since $f_o(S_{P_1})$ is a non-constant convex function, there will be only one possible value of $S_{P_1}$ for which its derivative equals 0. In particular, this means that the derivative cannot be equal to 0 at $S_{P_1}^{\min}$ and $S_{P_1}^{\max}$ simultaneously and, consequently, no ambiguity can happen when checking the optimality conditions of such extreme values.

In case that these extreme values are not optimum, then the optimum value $S_{P_1}^{\star}$ will be computed resorting to steps 17-22 in Table \ref{alg}. These steps allows to calculate numerically the value of $S_{P_1}$ within the interval $\left(S_{P_1}^{\min},S_{P_1}^{\max}\right)$ for which the derivative is 0. This is carried out by deriving successive nested intervals over variable $S_{P_1}$, each one with a length equal to one half of the length of the previous interval. The left extreme of the intervals is selected such that the derivative of $f_o$ is non-positive at such extreme (step 18), whereas the derivative is non-negative on the right extreme (step 19). Asymptotically, the length of the nested intervals tends to 0 and the central point of the interval tends to the optimum solution, i.e., the value of $S_{P_1}$ for which the derivative of $f_o(S_{P_1})$ equals 0 (step 22).

The derivatives of the convex function $f_o(S_{P_1})$ that appear in the previous paragraph and that are also used in the iterations based on the nested intervals approach detailed in Table \ref{alg} can be calculated according to the following expressions:
\begin{equation}
\frac{df_o(S_{P_1})}{dS_{P_1}}=\beta_{\mbox{\tiny{UL}}}\overline{e}_{\mbox{\tiny{UL}}}(r_{\mbox{\tiny{UL}}}^{\star}(S_{P_1}))+S_{P_1}\beta_{\mbox{\tiny{UL}}}\overline{e}_{\mbox{\tiny{UL}}}'(r_{\mbox{\tiny{UL}}}^{\star}(S_{P_1}))\frac{dr_{\mbox{\tiny{UL}}}^{\star}(S_{P_1})}{dS_{P_1}}+k_{\mbox{\scriptsize{rx}},1}\frac{\beta_{\mbox{\tiny{DL}}}}{R_{\mbox{\tiny{DL}}}^{\max}}+k_{\mbox{\scriptsize{rx}},2}\beta_{\mbox{\tiny{DL}}}-\varepsilon_{P_0},\label{eq:deriv_energy_wrt_sp1}
\end{equation}
where
\begin{eqnarray}
\overline{e}_{\mbox{\tiny{UL}}}'(r_{\mbox{\tiny{UL}}}) & = & -\frac{k_{\mbox{\scriptsize{tx}},1}}{r_{\mbox{\tiny{UL}}}^2}-\frac{k_{\mbox{\scriptsize{tx}},2}}{r_{\mbox{\tiny{UL}}}^2}\sum_{i=1}^{K(r_{\mbox{\tiny{UL}}})}\left(c(r_{\mbox{\tiny{UL}}})-\frac{1}{\lambda_i}\right) \nonumber \\ & & +\frac{k_{\mbox{\scriptsize{tx}},2}}{r_{\mbox{\tiny{UL}}}}\sum_{i=1}^{K(r_{\mbox{\tiny{UL}}})}\frac{\log2}{W_{\mbox{\tiny{UL}}}K(r_{\mbox{\tiny{UL}}})}\frac{2^{\frac{r_{\mbox{\tiny{UL}}}}{W_{\mbox{\tiny{UL}}}K(r_{\mbox{\tiny{UL}}})}}}{\left(\prod_{k=1}^{K(r_{\mbox{\tiny{UL}}})}\lambda_k\right)^{\frac{1}{K(r_{\mbox{\tiny{UL}}})}}},\\
\frac{dr_{\mbox{\tiny{UL}}}^{\star}(S_{P_1})}{dS_{P_1}} & = & \left\{\begin{array}[c]{ll} \frac{\beta_{\mbox{\tiny{UL}}}\frac{L_{\max}}{S_{P_1}^2}}{\left(\frac{L_{\max}}{S_{P_1}}-\tau_{P_1}-\frac{\beta_{\mbox{\tiny{DL}}}}{R_{\mbox{\tiny{DL}}}^{\max}}\right)^2}=\frac{\beta_{\mbox{\tiny{UL}}}L_{\max}}{\left(L_{\max}-S_{P_1}\tau_{P_1}-S_{P_1}\frac{\beta_{\mbox{\tiny{DL}}}}{R_{\mbox{\tiny{DL}}}^{\max}}\right)^2}, & r_{\mbox{\tiny{UL}}}^{\min}(S_{P_1})>\check{R}_{\mbox{\tiny{UL}}},\\
0, & r_{\mbox{\tiny{UL}}}^{\min}(S_{P_1})\leq \check{R}_{\mbox{\tiny{UL}}}.
\end{array}\right.
\end{eqnarray}
The previous expressions have been derived using the definition of function $\overline{e}_{\mbox{\tiny{UL}}}(r_{\mbox{\tiny{UL}}})$ provided in Lemma \ref{lemma:rate}, i.e., $\overline{e}_{\mbox{\tiny{UL}}}(r_{\mbox{\tiny{UL}}})=\overline{e}_{\mbox{\tiny{UL}}}\left(\frac{s_{\mbox{\tiny{UL}}}}{t_{\mbox{\tiny{UL}}}}\right)=\frac{1}{s_{\mbox{\tiny{UL}}}}e_{\mbox{\tiny{UL}}}(t_{\mbox{\tiny{UL}}},s_{\mbox{\tiny{UL}}})=e_{\mbox{\tiny{UL}}}\left(\frac{1}{r_{\mbox{\tiny{UL}}}},1\right)$, where the explicit expression of $e_{\mbox{\tiny{UL}}}(t_{\mbox{\tiny{UL}}},s_{\mbox{\tiny{UL}}})$ is given by (\ref{eq:ul_energy}).

\begin{table}[!t]
\renewcommand{\arraystretch}{1.1}
\caption{\textsc{Iterative Algorithm to Calculate the Optimum Value of
the Number of Bits to Be Transmitted through the UL in the Joint
Communication and Computation Resource Allocation Problem}}
\vspace{-0.6cm} \label{alg}
\begin{center}
\begin{tabular}{ l  l }
\hline \hline
1: & calculate $S_{P_1}^{\min}$ according to (\ref{eq:sp1_min}) \\
2: & calculate $S_{P_1}^{\max}$ according to (\ref{eq:sp1_max}) \\
3: & \textbf{if} $S_{P_1}^{\max}<S_{P_1}^{\min}$\\
4: & \quad \quad problem is infeasible: increase $L_{\max} \longrightarrow$ \textbf{go to} 24 \\
5: & \textbf{end if} \\
6: & calculate $\frac{df_o(S_{P_1}^{\min})}{dS_{P_1}}$ according to
(\ref{eq:deriv_energy_wrt_sp1})\\
7: & \textbf{if} $\frac{df_o(S_{P_1}^{\min})}{dS_{P_1}}\geq 0$\\
8: & \quad \quad $S_{P_1}^{\star}=S_{P_1}^{\min}$\\
9: & \quad \quad \textbf{go to} 23\\
10: & \textbf{end if} \\
11: & calculate $\frac{df_o(S_{P_1}^{\max})}{dS_{P_1}}$ according to
(\ref{eq:deriv_energy_wrt_sp1})\\
12: & \textbf{if} $\frac{df_o(S_{P_1}^{\max})}{dS_{P_1}}\leq 0$\\
13: & \quad \quad $S_{P_1}^{\star}=S_{P_1}^{\max}$\\
14: & \quad \quad \textbf{go to} 23\\
15: & \textbf{end if} \\
16: & set $S_{\inf}=S_{P_1}^{\min}$, $S_{\sup}=S_{P_1}^{\max}$, $\overline{S}=\frac{1}{2}\left(S_{\inf}+S_{\sup}\right)$\\
17: & \textbf{repeat}\\
18: & \quad \quad \textbf{if} $\frac{df_o(\overline{S})}{dS_{P_1}}\leq 0$, \textbf{then} set $S_{\inf}=\overline{S}$\\
19: & \quad \quad \textbf{otherwise}, set $S_{\sup}=\overline{S}$\\
20: & \quad \quad set $\overline{S}=\frac{1}{2}\left(S_{\inf}+S_{\sup}\right)$\\
21: & \textbf{until} $S_{\sup}-S_{\inf}<\epsilon\left(S_{P_1}^{\max}-S_{P_1}^{\min}\right)$\\
22: & take the last obtained value of $\overline{S}$ as a valid
approximation of the optimum solution: $S_{P_1}^{\star}\simeq
\overline{S}$\\
23: & based on $S_{P_1}^{\star}$, calculate the other parameters
involved in the problem: $S_{P_0}^{\star},r_{\mbox{\tiny{UL}}}^{\star},r_{\mbox{\tiny{DL}}}^{\star},t_{\mbox{\tiny{UL}}}^{\star},t_{\mbox{\tiny{DL}}}^{\star}$\\
24: & \textbf{end algorithm}\\

\hline \hline
\end{tabular}
\end{center}
\vspace{-0.8cm}
\end{table}

Summarizing, the main results of this section are the following. Given the application parameters (i.e., energy per processed bit required by the execution of the application at the MT, number input/output bits, etc.) we may minimize the energy consumption of the MT by optimizing the partition of the data to be processed locally and remotely and the UL transmission rate (as this will have an impact on the energy consumption of the MT when transmitting through the UL). We have found that for each possible partition there is an optimal transmission UL data rate which is given by eq. (\ref{eq:r_ul_star}). Then, we have proposed a method for calculating efficiently the optimal data partition in terms of the total energy consumption at the MT.

\section{Analysis of Particular Cases}\label{sec:particular_cases}
In this section we provide an analysis of a number of particular
cases of the general resource allocation problem defined and solved
in the previous sections. This analysis provides an
insight into the problem and the solution itself and give practical
guidelines for the application of the proposed strategy.

\subsection{Optimality of No Offloading}
In this subsection we provide the necessary and sufficient
conditions under which the optimum solution is to process all the
bits locally at the MT, i.e., $S_{P_1}^{\star}=0$. These conditions
are twofold: (i) $S_{P_1}=0$ should be feasible, and (ii)
$\frac{df_o(0)}{dS_{P_1}}\geq 0$.

According to (\ref{eq:sp1_min}), the first condition (i) holds if,
and only if, $L_{\max}\geq S_{\mbox{\scriptsize{app}}}\tau_{P_0}$, i.e., executing all the application locally at the MT does not violate the latency constraint.

On the other hand, we see from
(\ref{eq:global_optim_problem_simple}) that function
$r_{\mbox{\tiny{UL}}}^{\min}(S_{P_1})$ is equal to 0 at $S_{P_1}=0$ (i.e.,
$r_{\mbox{\tiny{UL}}}^{\min}(0)=0$) and is continuous within a certain interval
containing $S_{P_1}=0$. These two characteristics allow to state
that function $r_{\mbox{\tiny{UL}}}^{\star}(S_{P_1})$ will be constant (i.e., not
depending on $S_{P_1}$) also within a certain interval containing
$S_{P_1}=0$. The main consequence from this is that
$\frac{dr_{\mbox{\tiny{UL}}}^{\star}(0)}{dS_{P_1}}=0$ and, therefore, the second condition (ii) holds if, and only if,
\begin{equation}
\frac{df_o(0)}{dS_{P_1}}=\beta_{\mbox{\tiny{UL}}}\overline{e}_{\mbox{\tiny{UL}}}(r_{\mbox{\tiny{UL}}}^{\star}(0))+k_{\mbox{\scriptsize{rx}},1}\frac{\beta_{\mbox{\tiny{DL}}}}{R_{\mbox{\tiny{DL}}}^{\max}}+k_{\mbox{\scriptsize{rx}},2}\beta_{\mbox{\tiny{DL}}}-\varepsilon_{P_0}\geq 0.\label{eq:derivative_f0_0_no_offloading}
\end{equation}
The previous condition is equivalent to $\varepsilon_{P_0} \leq \beta_{\mbox{\tiny{UL}}}\overline{e}_{\mbox{\tiny{UL}}}(r_{\mbox{\tiny{UL}}}^{\star}(0))+k_{\mbox{\scriptsize{rx}},1}\frac{\beta_{\mbox{\tiny{DL}}}}{R_{\mbox{\tiny{DL}}}^{\max}}+k_{\mbox{\scriptsize{rx}},2}\beta_{\mbox{\tiny{DL}}}$, i.e., the energy required to process 1 bit locally at the MT ($\varepsilon_{P_0}$) should be lower than the energy required to transmit 1 bit through the UL ($\beta_{\mbox{\tiny{UL}}}\overline{e}_{\mbox{\tiny{UL}}}(r_{\mbox{\tiny{UL}}}^{\star}(0))$) plus the energy required to receive through the DL the output data portion corresponding to the processing of 1 bit ($k_{\mbox{\scriptsize{rx}},1}\frac{\beta_{\mbox{\tiny{DL}}}}{R_{\mbox{\tiny{DL}}}^{\max}}+k_{\mbox{\scriptsize{rx}},2}\beta_{\mbox{\tiny{DL}}}$). Note that if the channel conditions improve, then the terms $\beta_{\mbox{\tiny{UL}}}\overline{e}_{\mbox{\tiny{UL}}}(r_{\mbox{\tiny{UL}}}^{\star}(0))$ and $k_{\mbox{\scriptsize{rx}},1}\frac{\beta_{\mbox{\tiny{DL}}}}{R_{\mbox{\tiny{DL}}}^{\max}}$ would decrease and, therefore, total processing at the MT may not be optimum any more.

\subsection{Optimality of Total Offloading}
In this subsection we provide the necessary and sufficient
conditions under which the optimum solution is to process all the
bits remotely at the FAP, i.e., $S_{P_1}^{\star}=S_{\mbox{\scriptsize{app}}}$. These
conditions are twofold: (i) $S_{P_1}=S_{\mbox{\scriptsize{app}}}$ should be feasible,
and (ii) $\frac{df_o(S_{\mbox{\scriptsize{app}}})}{dS_{P_1}}\leq 0$.

According to (\ref{eq:sp1_max}), the first condition (i) holds if,
and only if, $L_{\max}\geq
S_{\mbox{\scriptsize{app}}}\left(\frac{\beta_{\mbox{\tiny{UL}}}}{R_{\mbox{\tiny{UL}}}^{\max}}+\tau_{P_1}+\frac{\beta_{\mbox{\tiny{DL}}}}{R_{\mbox{\tiny{DL}}}^{\max}}\right)$, i.e., the time required to transmit all the data through the UL, for the remote processing, and for the DL transmission of the output data, should not violate the maximum latency constraint.

Finally, the necessary and sufficient condition (ii) can be expanded as
\begin{equation}
\frac{df_o(S_{\mbox{\scriptsize{app}}})}{dS_{P_1}}=\beta_{\mbox{\tiny{UL}}}\overline{e}_{\mbox{\tiny{UL}}}(r_{\mbox{\tiny{UL}}}^{\star}(S_{\mbox{\scriptsize{app}}}))+S_{\mbox{\scriptsize{app}}}\beta_{\mbox{\tiny{UL}}}\overline{e}_{\mbox{\tiny{UL}}}'(r_{\mbox{\tiny{UL}}}^{\star}(S_{\mbox{\scriptsize{app}}}))\frac{dr_{\mbox{\tiny{UL}}}^{\star}(S_{\mbox{\scriptsize{app}}})}{dS_{P_1}}
+\,k_{\mbox{\scriptsize{rx}},1}\frac{\beta_{\mbox{\tiny{DL}}}}{R_{\mbox{\tiny{DL}}}^{\max}}+k_{\mbox{\scriptsize{rx}},2}\beta_{\mbox{\tiny{DL}}}-\varepsilon_{P_0}\leq
0.
\end{equation}

\subsection{Feasibility and Minimum Affordable Latency}
\begin{figure}[t]
    \begin{center}
        \includegraphics[width=11cm]{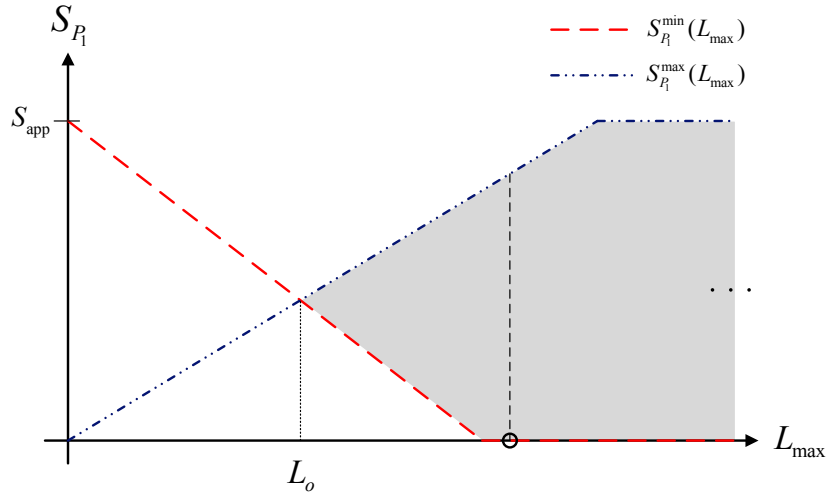}
    \end{center}
\vspace{-1cm}
    \caption{Dependency of $S_{P_1}^{\max}$ and $S_{P_1}^{\min}$ vs. $L_{\max}$.}
    \label{fig:sp1_max_min_lmax}
\end{figure}

As commented in the previous section, the problem
(\ref{eq:global_optim_problem}) to be solved is feasible if, and
only if, $S_{P_1}^{\max}\geq S_{P_1}^{\min}$. As
shown explicitly in (\ref{eq:sp1_min}) and (\ref{eq:sp1_max}), these
two values depend on the maximum allowed latency $L_{\max}$ (in
fact, they are linear functions of $L_{\max}$ with a top and a
bottom saturation at $S_{\mbox{\scriptsize{app}}}$ and $0$, respectively). The plot of
these two functions (i.e., $S_{P_1}^{\max}(L_{\max})$ and
$S_{P_1}^{\min}(L_{\max})$) is shown in Fig.
\ref{fig:sp1_max_min_lmax}. In such figure, each vertical segment
within the shaded region represents the set of feasible values of
$S_{P_1}$ for each value of $L_{\max}$ (see, as example, the dashed
vertical segment within the shaded region, that contains the
feasible values of $S_{P_1}$ for the corresponding value of
$L_{\max}$ represented in the figure through a circle).

From the figure it can be seen clearly that there will be a lowest
value of $L_{\max}$ (also called minimum affordable latency) under
which problem (\ref{eq:global_optim_problem}) becomes infeasible.
Let us denote such lowest value by $L_o$. Thanks to the closed-form
expressions (\ref{eq:sp1_min}) and (\ref{eq:sp1_max}), an analytic
expression for $L_o$ can be calculated (it is, in fact, the crossing
of the two-linear segments of functions $S_{P_1}^{\max}(L_{\max})$
and $S_{P_1}^{\min}(L_{\max})$). The minimum admissible value of the
latency for which the problem is feasible is, thus,
\begin{equation}
L_o=\frac{S_{\mbox{\scriptsize{app}}}}{\frac{1}{\frac{\beta_{\mbox{\tiny{UL}}}}{R_{\mbox{\tiny{UL}}}^{\max}}+\tau_{P_1}+\frac{\beta_{\mbox{\tiny{DL}}}}{R_{\mbox{\tiny{DL}}}^{\max}}}+\frac{1}{\tau_{P_0}}}=S_{\mbox{\scriptsize{app}}}\frac{\tau_{P_0}\left(\frac{\beta_{\mbox{\tiny{UL}}}}{R_{\mbox{\tiny{UL}}}^{\max}}+\tau_{P_1}+\frac{\beta_{\mbox{\tiny{DL}}}}{R_{\mbox{\tiny{DL}}}^{\max}}\right)}{\frac{\beta_{\mbox{\tiny{UL}}}}{R_{\mbox{\tiny{UL}}}^{\max}}+\tau_{P_1}+\frac{\beta_{\mbox{\tiny{DL}}}}{R_{\mbox{\tiny{DL}}}^{\max}}+\tau_{P_0}}.\label{eq:l_o}
\end{equation}
Note that $L_o$ is always lower than $S_{\mbox{\scriptsize{app}}}\tau_{P_0}$, i.e., the time that would be needed to do all the processing locally at the MT, and lower than $S_{\mbox{\scriptsize{app}}}\left(\frac{\beta_{\mbox{\tiny{UL}}}}{R_{\mbox{\tiny{UL}}}^{\max}}+\tau_{P_1}+\frac{\beta_{\mbox{\tiny{DL}}}}{R_{\mbox{\tiny{DL}}}^{\max}}\right)$, i.e., the time that would be needed to do all the processing remotely at the FAP (including UL transmission, processing, and DL transmission).

Interestingly, when the time budget (i.e., the maximum allowed latency $L_{\max}$) equals
the minimum affordable latency $L_o$, partial offloading is required, and the distribution of bits is given by
\begin{equation}
L_{\max}=L_o\;\;\;\Rightarrow\;\;\;\left\{\begin{array}[c]{l} S_{P_0}^{\star}=S_{\mbox{\scriptsize{app}}}\frac{\frac{\beta_{\mbox{\tiny{UL}}}}{R_{\mbox{\tiny{UL}}}^{\max}}+\tau_{P_1}+\frac{\beta_{\mbox{\tiny{DL}}}}{R_{\mbox{\tiny{DL}}}^{\max}}}{\frac{\beta_{\mbox{\tiny{UL}}}}{R_{\mbox{\tiny{UL}}}^{\max}}+\tau_{P_1}+\frac{\beta_{\mbox{\tiny{DL}}}}{R_{\mbox{\tiny{DL}}}^{\max}}+\tau_{P_0}} \\
S_{P_1}^{\star}=S_{\mbox{\scriptsize{app}}}\frac{\tau_{P_0}}{\frac{\beta_{\mbox{\tiny{UL}}}}{R_{\mbox{\tiny{UL}}}^{\max}}+\tau_{P_1}+\frac{\beta_{\mbox{\tiny{DL}}}}{R_{\mbox{\tiny{DL}}}^{\max}}+\tau_{P_0}}\end{array}\right.,\label{eq:optim_s_min_feasible_l}
\end{equation}
where the previous expressions have been obtained by calculating the
crossing point between $S_{P_1}^{\max}(L_{\max})$ and
$S_{P_1}^{\min}(L_{\max})$.

In some situations and for some concrete applications, the delay
experienced by the application is the only performance indicator that matters, while the energy spent by the MT does not play any important role. This can happen, for example, when we have a laptop or a smart-phone connected to the electric power grid (and, therefore, the battery is not a limitation) or when we are running an online interactive game where the latency should be as low as possible to perceive a real-time interaction among players. In these cases, the offloading design problem becomes the following:
\begin{equation}
\begin{array}
[c]{cl}
\operatorname*{minimize}\limits_{L,S_{P_0},S_{P_1},t_{\mbox{\tiny{UL}}},t_{\mbox{\tiny{DL}}}} & L \\
\operatorname*{subject\,\,to} &
\!\!\begin{array}
[t]{l} S_{P_0}+S_{P_1}=S_{\mbox{\scriptsize{app}}},\\
\max\left\{\tau_{P_0}S_{P_0},t_{\mbox{\tiny{UL}}}+\tau_{P_1}S_{P_1}+t_{\mbox{\tiny{DL}}}\right\}\leq
L,\\ e_{\mbox{\tiny{UL}}}(t_{\mbox{\tiny{UL}}},\beta_{\mbox{\tiny{UL}}}S_{P_1})-k_{\mbox{\scriptsize{tx}},1}t_{\mbox{\tiny{UL}}}\leq
k_{\mbox{\scriptsize{tx}},2}t_{\mbox{\tiny{UL}}}P_{\mbox{\scriptsize{tx}},\mbox{\tiny{MT}}},\\ \beta_{\mbox{\tiny{DL}}}S_{P_1}\leq
t_{\mbox{\tiny{DL}}}R_{\mbox{\tiny{DL}}}^{\max}.
\end{array}
\end{array}
\label{eq:minimum_latency_problem}
\end{equation}

The previous problem is, in fact, the feasibility test for the
original problem (\ref{eq:global_optim_problem}) and, therefore, the
optimum solution is given by $L^{\star}=L_o$ (\ref{eq:l_o}) and the
distribution of bits for processing detailed in
(\ref{eq:optim_s_min_feasible_l}).

\subsection{Minimum Energy without Latency Constraints}\label{subsec:min_energ_no_latency}
In situations where the MT has a very low battery level or for
applications which are delay-tolerant, the user may be interested in
minimizing the total energy spending no matter how much delay this
implies. In fact, this situation can be modeled using the general
problem formulation (\ref{eq:global_optim_problem}) but without
including constraint C2, or what is equivalent, by assuming that
$L_{\max}\rightarrow\infty$ (i.e., there is no effective latency constraint).

The solution to the previous problem can be found by just taking the
expressions for the general problem formulation and particularizing
them to the case of $L_{\max}\rightarrow\infty$. The first main conclusion is
that, according to (\ref{eq:sp1_min}-\ref{eq:sp1_max}), the feasible
set for variable $S_{P_1}$ is
\begin{equation}
0\leq S_{P_1} \leq S_{\mbox{\scriptsize{app}}}.
\end{equation}

We have also that, according to
(\ref{eq:global_optim_problem_simple}),
\begin{equation}
r_{\mbox{\tiny{UL}}}^{\min}(S_{P_1})=0,\;\;\;\forall S_{P_1}\in [0,S_{\mbox{\scriptsize{app}}}].
\end{equation}

Based on the previous result and using (\ref{eq:r_ul_star}), we
deduce that function $r_{\mbox{\tiny{UL}}}^{\star}(S_{P_1})$ is constant and,
therefore, denoted in what follows simply by $r_{\mbox{\tiny{UL}}}^{\star}$ (with
a value equal to either $\check{R}_{\mbox{\tiny{UL}}}$ or $R_{\mbox{\tiny{UL}}}^{\max}$), which
implies that the derivative of $r_{\mbox{\tiny{UL}}}^{\star}(S_{P_1})$ w.r.t. $S_{P_1}$ is zero:
\begin{equation}
r_{\mbox{\tiny{UL}}}^{\star}(S_{P_1})=r_{\mbox{\tiny{UL}}}^{\star}=\left\{\begin{array}[c]{ll}
\check{R}_{\mbox{\tiny{UL}}}, & \check{R}_{\mbox{\tiny{UL}}}\leq R_{\mbox{\tiny{UL}}}^{\max},
\\ R_{\mbox{\tiny{UL}}}^{\max},& \check{R}_{\mbox{\tiny{UL}}}
> R_{\mbox{\tiny{UL}}}^{\max}
\end{array}\right.\;\;\;\Rightarrow\;\;\;\frac{dr_{\mbox{\tiny{UL}}}^{\star}(S_{P_1})}{dS_{P_1}}=0,\;\;\forall S_{P_1}\in [0,S_{\mbox{\scriptsize{app}}}].
\end{equation}

Finally, by collecting all the previous results, the total energy
spending (\ref{eq:deriv_energy_wrt_sp1}) can be rewritten as
\begin{equation}
f_o(S_{P_1})=\left(\beta_{\mbox{\tiny{UL}}}\overline{e}_{\mbox{\tiny{UL}}}(r_{\mbox{\tiny{UL}}}^{\star})+k_{\mbox{\scriptsize{rx}},1}\frac{\beta_{\mbox{\tiny{DL}}}}{R_{\mbox{\tiny{DL}}}^{\max}}+k_{\mbox{\scriptsize{rx}},2}\beta_{\mbox{\tiny{DL}}}-\varepsilon_{P_0}\right)S_{P_1}+\varepsilon_{P_0}S_{\mbox{\scriptsize{app}}},
\end{equation}
from which it is seen that, in this case, the dependency of the
energy with $S_{P_1}$ is linear. Based on this, we find the
optimum solution to the problem as
\begin{equation}
S_{P_1}^{\star}=\left\{\begin{array}[c]{ll} 0, &\;\;\;
\beta_{\mbox{\tiny{UL}}}\overline{e}_{\mbox{\tiny{UL}}}(r_{\mbox{\tiny{UL}}}^{\star})+k_{\mbox{\scriptsize{rx}},1}\frac{\beta_{\mbox{\tiny{DL}}}}{R_{\mbox{\tiny{DL}}}^{\max}}+k_{\mbox{\scriptsize{rx}},2}\beta_{\mbox{\tiny{DL}}} \geq \varepsilon_{P_0},
\\ S_{\mbox{\scriptsize{app}}},&\;\;\;
\beta_{\mbox{\tiny{UL}}}\overline{e}_{\mbox{\tiny{UL}}}(r_{\mbox{\tiny{UL}}}^{\star})+k_{\mbox{\scriptsize{rx}},1}\frac{\beta_{\mbox{\tiny{DL}}}}{R_{\mbox{\tiny{DL}}}^{\max}}+k_{\mbox{\scriptsize{rx}},2}\beta_{\mbox{\tiny{DL}}} < \varepsilon_{P_0}.
\end{array}\right.\label{eq:min_energ_no_latency_const}
\end{equation}
From the previous result it is concluded that, without latency constraint, partial offloading can never be optimal, i.e., the optimum solution in terms of energy consumption is to process all the data either locally or remotely. In this situation, we would like to emphasize that from (\ref{eq:min_energ_no_latency_const}) we can find the optimal decision basically from the comparison of the energy that would be needed to process 1 bit locally (represented by $\varepsilon_{P_0}$) and the energy that would be required to transmit 1 bit through the UL, to process such bit remotely at the FAP, and to send the corresponding output data to the MT through the DL (represented by $\beta_{\mbox{\tiny{UL}}}\overline{e}_{\mbox{\tiny{UL}}}(r_{\mbox{\tiny{UL}}}^{\star})+k_{\mbox{\scriptsize{rx}},1}\frac{\beta_{\mbox{\tiny{DL}}}}{R_{\mbox{\tiny{DL}}}^{\max}}+k_{\mbox{\scriptsize{rx}},2}\beta_{\mbox{\tiny{DL}}}$). Note that if the channel condition improves, then the terms $\beta_{\mbox{\tiny{UL}}}\overline{e}_{\mbox{\tiny{UL}}}(r_{\mbox{\tiny{UL}}}^{\star})$ and $k_{\mbox{\scriptsize{rx}},1}\frac{\beta_{\mbox{\tiny{DL}}}}{R_{\mbox{\tiny{DL}}}^{\max}}$ would decrease and, therefore, it would be more likely that the optimum solution is total offloading.

\subsection{Summary}
Summarizing, the main results of this section are the following. Given a certain set of parameters and channel conditions, if the problem is not latency-constrained (that is, if the latency constraint C2 is problem (\ref{eq:global_optim_problem}) is not fulfilled with equality), then the optimal solution in terms of total energy consumed by the MT is to do all the processing either locally or remotely. In such a situation, when offloading is optimum, the optimal UL data rate is the one minimizing the energy consumption per bit. In case that the system is constrained by the maximum affordable latency, the optimal UL data rate depends on the concrete partition considered. Still in such a situation, conditions for which total offloading or no offloading are optimum have been found. Finally, if the goal if to minimize the latency, then partial offloading is required and the optimum partition depends on the maximum UL and DL data rates possible according to the power budget for both the MT and the FAP.

\section{Simulation Results}\label{sec:simulations}
This section provides some simulations results to illustrate the performance of the proposed offloading optimization strategy. In all the presented simulations, the following numerical values for the parameters related with the energy consumption model in (\ref{eq:ul_power_model}) and (\ref{eq:dl_power_model}) have been taken: $k_{\mbox{\scriptsize{tx}},1}=0.4$ W, $k_{\mbox{\scriptsize{tx}},2}=18$, $k_{\mbox{\scriptsize{rx}},1}=0.4$ W, $k_{\mbox{\scriptsize{rx}},2}=2.86$ W/Mbps. These values have been computed through numerical regressions to be aligned with the experimental measurements provided in \cite{jensen:12} for a LTE-MT dongle which, in turn, validates the power consumption models proposed by the European EARTH project \cite{earth_d23} and allows us to obtain realistic simulations results. To evaluate the actual impact of the offloading on the energy consumption, $k_{\mbox{\scriptsize{tx}},1}$ does not include the base power consumption measured without scheduled traffic, but only the base power increase for having the transmitter and receiver chains active with scheduled traffic.

Other physical parameters related to the channel bandwidth and the maximum radiated powers for the MT and the FAP that have been used in the simulations are: $W_{\mbox{\tiny{UL}}}=W_{\mbox{\tiny{DL}}}=10$ MHz, and $P_{\mbox{\scriptsize{tx}},\mbox{\tiny{MT}}}=P_{\mbox{\scriptsize{tx}},\mbox{\tiny{FAP}}}=100$ mW. In the simulations, unless stated otherwise, we have taken as the maximum allowed latency the value $L_{\max}=4$ s.

In \cite{miettinen:10}, the speed and computational energy characteristics of two mobile devices, Nokia N810 and N900, were provided. According to Table 1 in \cite{miettinen:10}, we will consider in our simulations the N810 device with an energy consumption of $\frac{1}{480\cdot 10^6}$ J/cycle when working at a CPU rate of $400\cdot 10^6$ cycles/s. The same paper provides the relation between the number of computational cycles and input bits for several applications. In particular, for the gzip compression application (that we will consider in our simulations), this number is 330 cycles/byte according to Table 3 in \cite{miettinen:10}. From these quantities, we can calculate the time required to process 1 bit ($\tau_{P_0}=10^{-7}$ s/bit) and the energy spent in the processing of 1 bit ($\varepsilon_{P_0}=8.6\cdot10^{-8}$ J/bit). As mentioned before, we will consider a gzip application compressing a set of files with a total size equal to $S_{\mbox{\scriptsize{app}}}=5$ MBytes, $\beta_{\mbox{\tiny{UL}}}=1$, $\beta_{\mbox{\tiny{DL}}}=0.2$ (note that we are considering that the compression application is able to generate output files with a size equal to 20\% of sizes of the input files before compression). Concerning the speed of the CPU at the FAP, we will assume that it is twice faster, which translates into $\tau_{P_1}=\tau_{P_0}/2$ (this can be achieved using a different processor or two processors in parallel with the same capabilities).

\begin{figure}[t]
    \begin{center}
        \includegraphics[width=11.5cm]{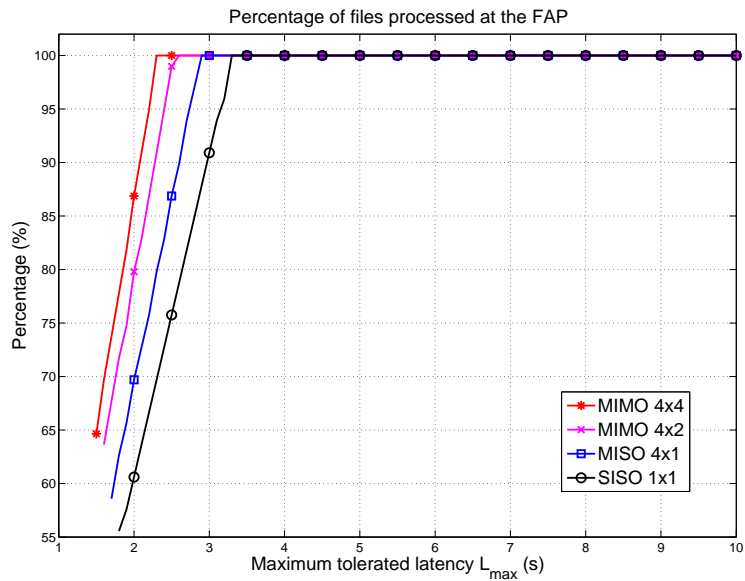}
    \end{center}
\vspace{-1.2cm}
    \caption{Percentage of files processed remotely at the FAP vs. maximum allowed latency $L_{\max}$ for a single channel realization.}
    \label{fig:files_latency}
\end{figure}

\begin{figure}[t]
    \begin{center}
        \includegraphics[width=11.5cm]{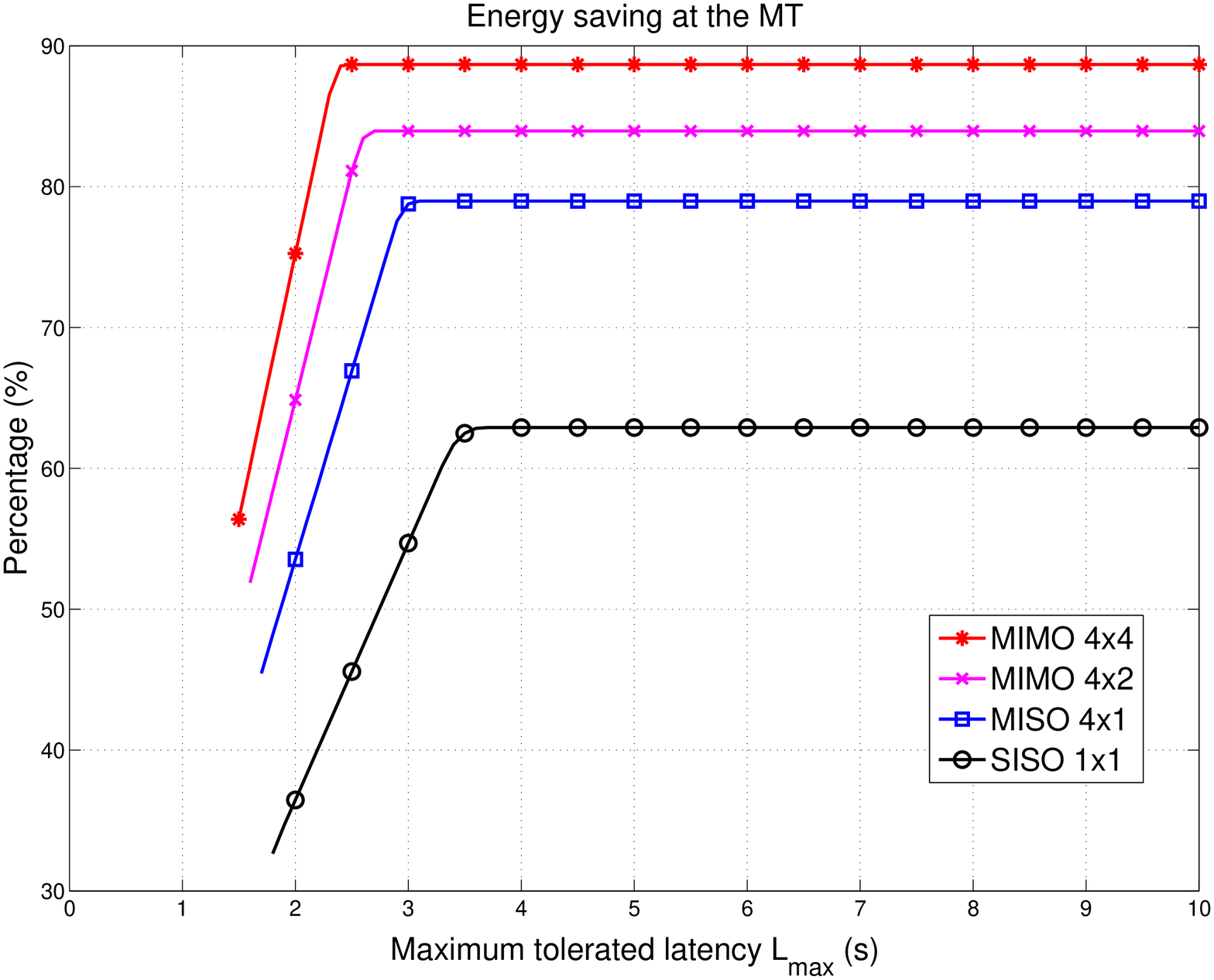}
    \end{center}
\vspace{-1.2cm}
    \caption{Percentage of energy saving thanks to offloading vs. maximum allowed latency $L_{\max}$ for a single channel realization.}
    \label{fig:energy_saving_latency}
\end{figure}

In the simulations we have considered four different cases of number of antennas: MIMO 4x4, MIMO 4x2, MISO 4x1, and SISO 1x1. Each point in the curves has been obtained by averaging 1000 random channels (except in Fig. \ref{fig:files_latency} and \ref{fig:energy_saving_latency}), where the channel matrix realization for each of them has been obtained by generating i.i.d. random zero-mean complex circularly symmetric Gaussian components with a variance equal to 1. In the figures, the behaviour of the system is analyzed as a function of $\gamma$. This parameter represents the mean channel gain (the same in UL and DL) normalized by the noise power and corresponds to a scalar factor multiplying the randomly generated channel matrices.

\begin{figure}[t]
    \begin{center}
        \includegraphics[width=11.5cm]{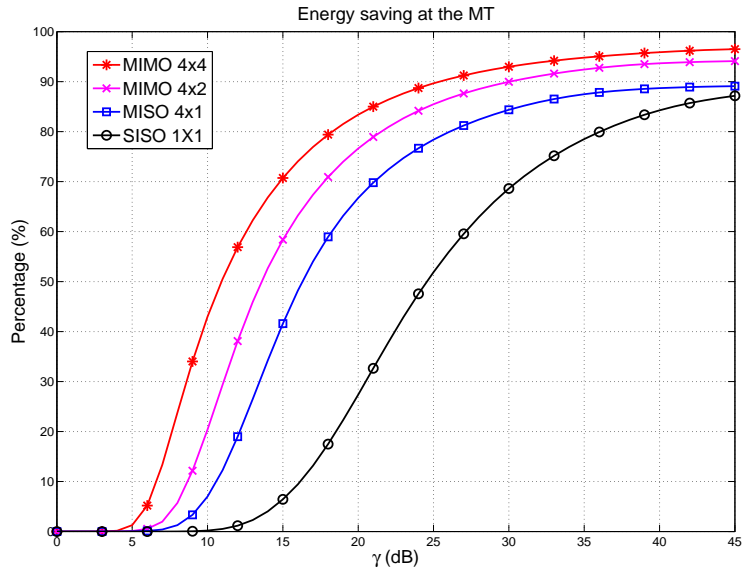}
    \end{center}
\vspace{-1.2cm}
    \caption{Percentage of energy saving thanks to offloading vs. mean channel gain $\gamma$.}
    \label{fig:energy_saving}
\end{figure}

Fig. \ref{fig:energy_saving} shows the energy saving in percentage w.r.t. the case of no offloading. Note that in the case of no offloading, the total energy spent would be $\varepsilon_{P_0}S_{\mbox{\scriptsize{app}}}$. On the other hand, the actual spent energy corresponds to the optimum solution of problem (\ref{eq:global_optim_problem}). As shown in the figure, when the number of antennas increases and the channel gain increases, the percentage of energy saving also improves, as expected.

\begin{figure}[t]
    \begin{center}
        \includegraphics[width=11.5cm]{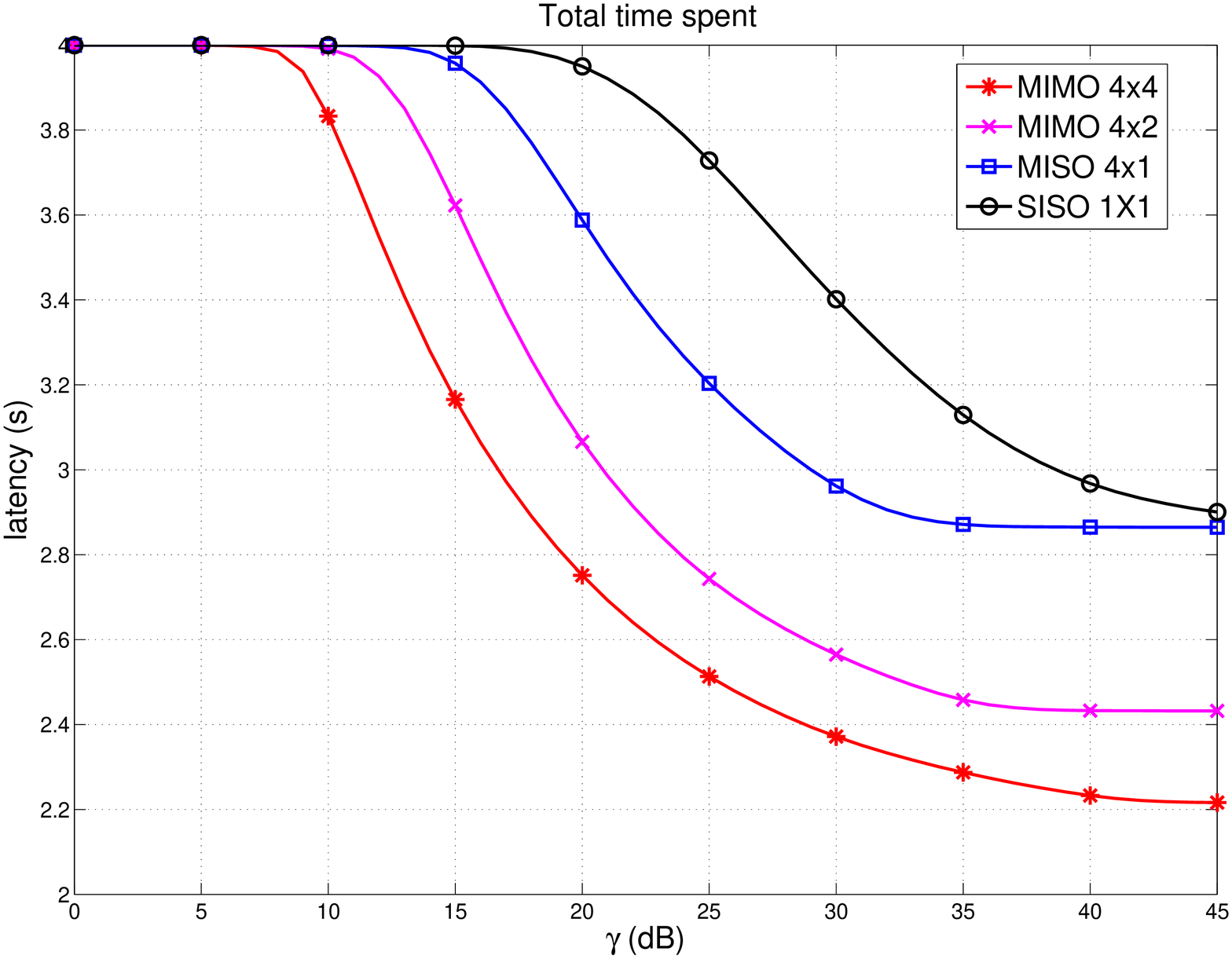}
    \end{center}
\vspace{-1.2cm}
    \caption{Actual latency vs. mean channel gain $\gamma$.}
    \label{fig:time_spent}
\end{figure}

Although the maximum latency constraint is set to $L_{\max}=4$ s, in some cases it may happen that constraint C2 in (\ref{eq:global_optim_problem}) is not fulfilled with equality in the optimum solution, i.e., in some situations the minimum energy spending is obtained with a latency even lower than the available time budget $L_{\max}$. We can see this effect in Fig. \ref{fig:time_spent}, that evaluates numerically the mean value of the actual latency as a function of the mean channel gain $\gamma$ for different antenna configurations. Note that, for low values of $\gamma$, all the allowed latency is used, but for higher values of the channel gain, the actual spent time decreases below $L_{\max}$. Basically, this happens due to the non-negligible term $k_{\mbox{\scriptsize{tx}},1}$ in the UL transmission power model (\ref{eq:ul_power_model}). Note that, as shown in Fig. \ref{fig:N_eUL_vs_rate}, at some point there is no energy saving in the UL communication from reducing the UL transmission rate (i.e., increasing the UL transmission time). This will happen whenever the optimum UL rate $r_{\mbox{\tiny{UL}}}^{\star}(S_{P_1}^{\star})$ is either $\check{R}_{\mbox{\tiny{UL}}}$ or $R_{\mbox{\tiny{UL}}}^{\max}$ (which is equivalent to the condition $\check{R}_{\mbox{\tiny{UL}}}\geq r_{\mbox{\tiny{UL}}}^{\min}(S_{P_1}^{\star})$, according to (\ref{eq:r_ul_star})).

\begin{figure}[t]
    \begin{center}
        \includegraphics[width=11.5cm]{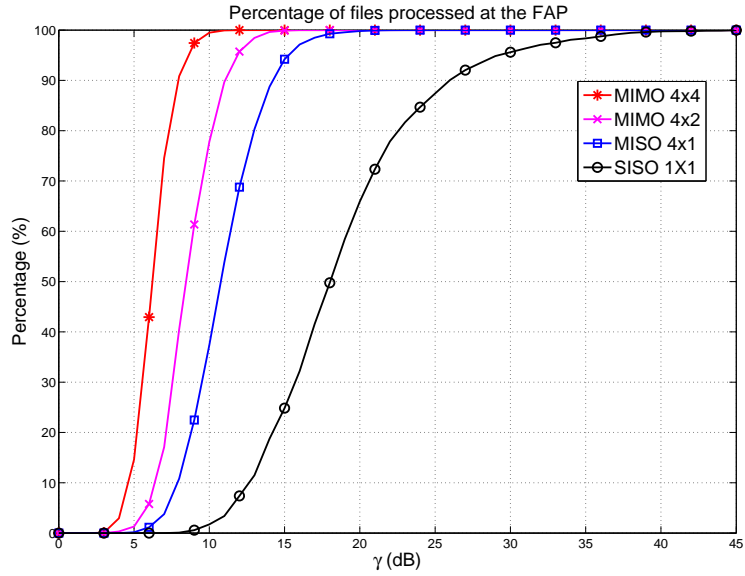}
    \end{center}
\vspace{-1.2cm}
    \caption{Percentage of files processed remotely at the FAP vs. mean channel gain $\gamma$.}
    \label{fig:file_processed}
\end{figure}

Although the previous two figures show the actual evaluation of the system performance in terms of the inherent relationship between energy and latency, it is important to get some insight into the actual system behaviour. In that sense, Fig. \ref{fig:file_processed} shows the percentage of the files processed remotely, i.e., $S_{P_1}^{\star}/S_{\mbox{\scriptsize{app}}}$, as a function of the mean channel gain $\gamma$. For very low values of the channel gain, sending the data through the communication channel would be very costly in terms of energy and, therefore, all the files are processed locally at the MT. As the channel gain increases, the percentage of files processed remotely also increases, arriving to total offloading at high channel gains. As expected, as the number of antennas increases, more bits will be processed remotely.

\begin{figure}[t]
    \begin{center}
        \includegraphics[width=11.5cm]{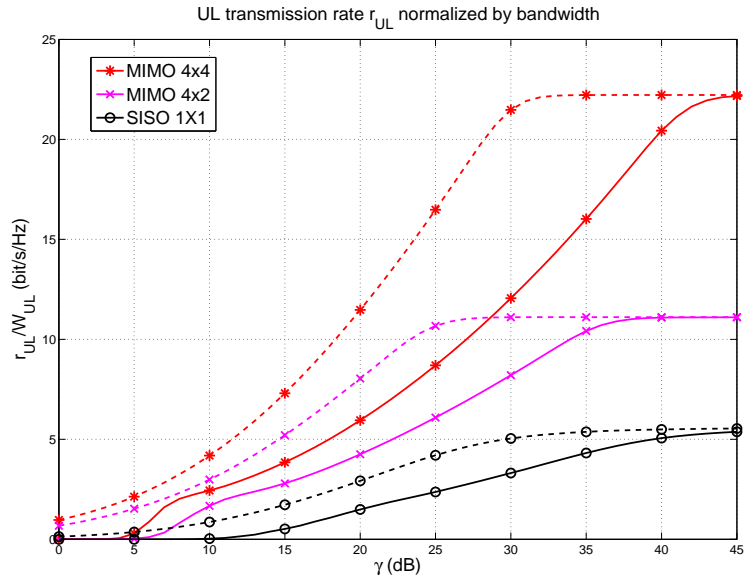}
    \end{center}
\vspace{-1.2cm}
    \caption{UL data rate $r_{\mbox{\tiny{UL}}}$ normalized by bandwidth vs. mean channel gain $\gamma$.}
    \label{fig:MT_rate}
\end{figure}

In Fig. \ref{fig:MT_rate} we show the UL data rate (in bits/s/Hz) corresponding to the optimum solution. It can be observed that as the channel quality increases, the data UL rate also increases. In the figure, the dashed line represents the maximum data rate $R_{\mbox{\tiny{UL}}}^{\max}$ allowed by the channel (see Eqs. (\ref{eq:reformulated_maximum_ul_rate})-(\ref{eq:reformulated_maximum_ul_rate2})). On the other hand, the solid line represents the actual UL rate resulting from the solution of problem in (\ref{eq:global_optim_problem}) which, in general, will be lower than $R_{\mbox{\tiny{UL}}}^{\max}$. As can be seen, both curves saturate at high values of $\gamma$. This happens because we have included an additional constraint concerning the maximum rate coming from practical aspects derived from the standard. In particular, we have set a maximum rate of 5.5 bit/s/Hz (maximum modulation and coding scheme allowed in LTE \cite{sesia:09}), multiplied by the maximum possible number of eigenmodes, which is equal to $\min\{n_{\mbox{\tiny{MT}}},n_{\mbox{\tiny{FAP}}}\}$. This is the constraint that has been added to the previous simulations and that generates the saturation effect in Fig. \ref{fig:MT_rate}.

Finally, in Fig. \ref{fig:files_latency} and \ref{fig:energy_saving_latency} we consider a single channel realization taken from a Rayleigh distribution with a mean channel gain of 25 dB. This approach is taken in order to understand better the impact of the latency constraint on the offloading process. Fig. \ref{fig:files_latency} shows the percentage of files to be processed remotely at the FAP as a function of the maximum allowed latency $L_{\max}$. Note that for a tight latency constraint, partial offloading is needed. On the other hand, and according to the results obtained in subsection \ref{subsec:min_energ_no_latency}, when the maximum tolerated latency is very high, the optimum solution is to perform the processing of all the files either locally at the MT or remotely at the FAP (see the conditions in (\ref{eq:min_energ_no_latency_const})). In the concrete case of the channel realization considered in this figure, the optimum solution for very high tolerated latencies is to offload all the files for all the considered cases in terms of number of antennas. Finally, Fig. \ref{fig:energy_saving_latency} shows the percentage of energy saving achieved from the offloading under the same conditions as in Fig. \ref{fig:files_latency}. We observe that relaxing the latency constraint allows for better energy savings. Note also that the energy saving saturates as from a certain value of the latency constraint (and beyond) the UL data rate for minimum energy can be afforded.

\section{Conclusions and Future Work}\label{sec:conclusions}
This paper has presented a general framework to optimize the communication and computational
resources usage in a scenario where an energy-limited MT in a femto-cell network intends
to run a computationally demanding application. In this framework, a
decision has to be taken regarding whether it is beneficial or not
to (partially) offload the application to the serving FAP. A
theoretical formulation of the problem has been presented
and solved providing some closed-form expressions that allow
simplifying significantly the optimization and the understanding of
the inherent tradeoff between energy spending and latency in both
the communication and computation stages. Finally, some particular
cases derived from the general design problem have been analyzed to
further understand the problem.

Although this paper has presented a general framework, it is
important to emphasize that the proposed solution is applicable to data-partitioned oriented applications with a predefined
amount of data to be processed. In addition, it has been assumed that the pool of bits to be processed can be divided between local
and remote processing without constraints related to the sizes of the two groups resulting from the data partitioning. Further work is to be
done to extend this approach to the case of applications with modularity constraints or with a-priori
predefined execution structure. Concerning the remote execution,
the possibility of allowing multiple FAPs to execute in parallel the modules of the
application is still to be analyzed. In relation with the communication, possible extensions
could include, for example, cooperative transmission schemes.
Finally, another possible future research line would consist in
extending the proposed strategy to the multiuser scenario, where
the available radio-communication and computational resources should be allocated using a proper scheduling strategy as a function of the
QoS demands and the channel states.

As mentioned in Section \ref{sec:scenario_problem}, the proposed offloading strategy is valid when the channel remains constant during the whole offloading process, which fits some realistic scenarios. In the case that the users have a mobility such that the previous assumption is not valid, the following two options could be considered to adapt our algorithm to the case of time-varying channels taking into account that the offloading decision has to be taken based only on a causal knowledge of the channel state:
\begin{itemize}
\item \emph{Suboptimum approach:} the complete set of data could be divided into smaller subsets. For each of these subsets an offloading decision should be taken taking into account the channel state at that moment. If the data subsets are small enough, it can be assumed that the channel remains constant during the potential offloading of each subset. This is a suboptimum approach since the optimum solution would take a global decision for all the subsets jointly, although this is no possible due to the fact that future channel states cannot be known in advance.
\item \emph{Optimum statistical approach:} problem (\ref{eq:global_optim_problem}) could be reformulated so that both the objective function and the constraints are replaced by the average expressions (or, alternatively, by the outage expressions) with respect to the channel statistics. This would allow taking a statistical offloading decision that would change if the channel statistics changes but that does not depend on the instantaneous channel state. However, finding a closed form solution or simple algorithm to obtain the optimum solution to this average formulation is quite complicated. A possible (and simpler) approach would consist in applying the philosophy presented in \cite{ribeiro:10}, which proposes an instantaneous stochastic gradient search algorithm to deal with this kind of problems.
\end{itemize}
The details and analysis of the previous approaches are out of the scope of this paper and are left for future research.

\bibliographystyle{IEEEtran}

\bibliography{IEEEabrv,references}

\begin{thebibliography}{10}
\providecommand{\url}[1]{#1}
\csname url@samestyle\endcsname
\providecommand{\newblock}{\relax}
\providecommand{\bibinfo}[2]{#2}
\providecommand{\BIBentrySTDinterwordspacing}{\spaceskip=0pt\relax}
\providecommand{\BIBentryALTinterwordstretchfactor}{4}
\providecommand{\BIBentryALTinterwordspacing}{\spaceskip=\fontdimen2\font plus
\BIBentryALTinterwordstretchfactor\fontdimen3\font minus
  \fontdimen4\font\relax}
\providecommand{\BIBforeignlanguage}[2]{{%
\expandafter\ifx\csname l@#1\endcsname\relax
\typeout{** WARNING: IEEEtran.bst: No hyphenation pattern has been}%
\typeout{** loaded for the language `#1'. Using the pattern for}%
\typeout{** the default language instead.}%
\else
\language=\csname l@#1\endcsname
\fi
#2}}
\providecommand{\BIBdecl}{\relax}
\BIBdecl

\bibitem{chandrasekhar:08}
V.~Chandrasekhar and J.~Andrews, ``{Femtocell Networks: A Survey},''
  \emph{{IEEE} Commun. Mag.}, vol.~46, no.~9, pp. 59--67, Sept. 2008.

\bibitem{luening:09}
J.~Luening and J.~Randolph, ``{Femtocells Economics},'' Mobile World Conference
  Barcelona (Barcelona), Febr. 2009.

\bibitem{zhu:11}
W.~Zhu, C.~Luo, J.~Wang, and S.~Li, ``{Multimedia Cloud Computing. An Emerging
  Technology for Providing Multimedia Services and Applications]},''
  \emph{{IEEE} Signal Process. Mag.}, vol.~28, no.~3, pp. 59--69, May 2011.

\bibitem{lei:13}
L.~Lei, Z.~Zhong, K.~Zheng, J.~Chen, and H.~Meng, ``{Challenges on Wireless
  Heterogeneous Networks for Mobile Cloud Computing},'' \emph{{IEEE} Wireless
  Commun. Mag.}, vol.~20, no.~3, pp. 34--44, June 2013.

\bibitem{gkatzikis:13}
L.~Gkatzikis and I.~Koutsopoulos, ``{Migrate or Not? Exploiting Dynamic Task
  Migration in Mobile Cloud Computing Systems},'' \emph{{IEEE} Wireless Commun.
  Mag.}, vol.~20, no.~3, pp. 24--32, June 2013.

\bibitem{miettinen:10}
A.~Miettinen and J.~Nurminen, ``{Energy Efficiency of Mobile Clients in Cloud
  Computing},'' in \emph{Proc. 2nd USENIX Conference on Hot Topics in Clod
  Computing 2010 (HotClout'10)}, June 2010.

\bibitem{kosta:12}
S.~Kosta, A.~Aucinas, P.~Hui, R.~Mortier, and X.~Zhang, ``{Thinkair: Dynamic
  Resource Allocation and Parallel Execution in the Cloud for Mobile Code
  Offloading},'' in \emph{Proc. IEEE International Conference on Computer
  Communications (INFOCOM'12)}, March 2012, pp. 945--953.

\bibitem{cuervo:10}
E.~Cuervo, A.~Balasubramanian, D.-K. Cho, A.~Wolman, S.~Saroiu, R.~Chandra, and
  P.~Bahl, ``{MAUI: Making Smartphones Last Longer with Code Offload},'' in
  \emph{Proc. International Conference on Mobile Systems, Applications, and
  Services (MobiSys'10)}, June 2010, pp. 49--62.

\bibitem{kumar:10}
K.~Kumar and Y.-H. Lu, ``{Cloud Computing for Mobile Users: Can Offloading
  Computation Save Energy?}'' \emph{{IEEE} Computer}, vol.~43, no.~4, pp.
  51--56, Apr. 2010.

\bibitem{kumar:13}
K.~Kumar, J.~Liu, Y.-H. Lu, and B.~Bhargava, ``{A Survey of Computation
  Of?oading for Mobile Systems},'' \emph{Mobile Networks and Applications,
  Springer Science}, vol.~18, no.~1, pp. 129--140, Febr. 2013.

\bibitem{lagerspetz:11}
E.~Lagerspetz and S.~Tarkoma, ``{Mobile Search and the Cloud: The Benefits of
  Offloading},'' in \emph{Proc. IEEE International Conference on Pervasive
  Computing and Communication (PerCom'11)}, March 2011, pp. 117--122.

\bibitem{kovachev:12}
D.~Kovachev and R.~Klamma, ``{Framework for Computation Offloading in Mobile
  Cloud Computing},'' \emph{International Jorunal of Interactive Multimedia and
  Artificial Intelligence}, vol.~1, no.~7, pp. 6--15, Dec. 2012.

\bibitem{zhang:13}
W.~Zhang, Y.~Wen, K.~Guan, D.~Kilper, H.~Luo, and D.~Wu, ``{Energy-Optimal
  Mobile Cloud Computing under Stochastic Wireless Channel},'' \emph{{IEEE}
  Trans. Wireless Commun.}, vol.~12, no.~9, pp. 4569--4581, Sept. 2013.

\bibitem{palomar:03}
D.~P. Palomar, J.~M. Cioffi, and M.~A. Lagunas, ``{Joint {Tx-Rx} Beamforming
  Design for Multicarrier {MIMO} Channels: A Unified Framework for Convex
  Optimization},'' \emph{{IEEE} Trans. Signal Process.}, vol.~51, no.~9, pp.
  2381--2401, Sept. 2003.

\bibitem{munoz:13}
O.~Muñoz, A.~Pascual-Iserte, and J.~Vidal, ``{Joint Allocation of Radio and
  Computational Resources in Wireless Application Offloading},'' in \emph{Proc.
  Future Network \& Mobile Summit (FUNEMS'13)}, July 2013.

\bibitem{zhang:10}
Q.~Zhang, L.~Cheng, and R.~Boutaba, ``{Cloud Computing: State-of-the-Art and
  Research Challenges},'' \emph{{Journal of Internet Services and
  Applications}}, vol.~1, no.~1, pp. 7--18, May 2010.

\bibitem{wang:13}
S.~Wang and S.~Dey, ``{Adaptive Mobile Cloud Computing to Enable Rich Mobile
  Multimedia Applications},'' \emph{{IEEE} Trans. Multimedia}, vol.~15, no.~4,
  pp. 870--883, June 2013.

\bibitem{molina:14}
M.~Molina, O.~Muñoz, A.~Pascual-Iserte, and J.~Vidal, ``{Joint Scheduling of
  Communication and Computation Resources in Multiuser Wireless Application
  Offloading},'' in \emph{Proc. IEEE International Symposium on Personal,
  Indoor and Mobile Radio Communications (PIMRC'14)}, Sept. 2014.

\bibitem{munoz:14}
O.~Muñoz, A.~Pascual-Iserte, J.~Vidal, and M.~Molina, ``{Energy-Latency
  Trade-off for Multiuser Wireless Computation Offloading},'' in \emph{Proc.
  IEEE Wireless Communications and Networking Conference (WCNC'14), workshop
  CLEEN (Workshop on Cloud Technologies and Energy Efficiency in Mobile
  Communication Networks)}, April 2014.

\bibitem{barbera:14}
M.~V. Barbera, S.~Kosta, A.~Mei, V.~C. Perta, and J.~Stefa, ``{Mobile
  Offloading in the Wild: Findings and Lessons Learned Through a Real-life
  Experiment with a New Cloud-aware System},'' in \emph{Proc. IEEE
  International Conference on Computer Communications (INFOCOM'14)}, April
  2014.

\bibitem{lorch:01}
J.~Lorch and J.~Smith, ``{Improving Dynamic Voltage Scaling Algorithms with
  PACE},'' in \emph{Proc. ACM SIGMETRICS Conference (SIGMETRICS'01)}, June
  2001, pp. 50--61.

\bibitem{sesia:09}
S.~Sesia, I.~Toufik, and M.~Baker, Eds., \emph{{LTE-The UMTS Long Term
  Evolution: From Theory to Practice}}.\hskip 1em plus 0.5em minus 0.4em\relax
  {John Wiley \& Sons}, 2009.

\bibitem{earth_d23}
G.~Auer, O.~Blume, V.~Giannini, I.~Godor, M.~Ali~Imran, Y.~Jading,
  E.~Katranaras, M.~Olsson, D.~Sabella, P.~Skillermark, and W.~Wajda, ``{Energy
  Efficiency Analysis of the Reference Systems, Areas of Improvements and
  Target Breakdown},'' deliverable report D2.3, ICT-247733 EARTH project,
  available at: https://www.ict-earth.eu/, Tech. Rep., Jan. 2012.

\bibitem{jensen:12}
A.~Jensen, M.~Lauridsen, P.~E. Mogensen, T.~Sørensen, and P.~Jensen, ``{LTE UE
  Power Consumption Model: For System Level Energy and Performance
  Optimization},'' in \emph{Proc. IEEE Vehicular Technology Conference Fall
  (VTC'12)}, Sept. 2012, pp. 1--5.

\bibitem{cui:03}
S.~Cui, A.~Goldsmith, and A.~Bahai, ``{Power Estimation for Viterbi
  Decoders},'' Wireless Systems Lab, Stanford Univ., CA, Tech. Rep. available
  at: http://wsl.stanford.edu/Publications.html, Tech. Rep., 2003.

\bibitem{raleigh:98}
G.~Raleigh and J.~Cioffi, ``{Spatio-Temporal Coding for Wireless
  Communication},'' \emph{{IEEE} Trans. Commun.}, vol.~46, no.~3, pp. 357--366,
  March 1998.

\bibitem{boyd:04}
S.~Boyd and L.~Vandenberghe, \emph{{Convex Optimization}}.\hskip 1em plus 0.5em
  minus 0.4em\relax {Cambridge University Press}, 2004.

\bibitem{quarteroni:07}
A.~Quarteroni, R.~Sacco, and F.~Saleri, Eds., \emph{{Numerical Mathematics
  (Texts in Applied Mathematics)}}, 2nd~ed.\hskip 1em plus 0.5em minus
  0.4em\relax {Springer-Verlag}, 2007.

\bibitem{suli:03}
E.~Süli and D.~Mayers, \emph{{An Introduction to Numerical Analysis}}.\hskip
  1em plus 0.5em minus 0.4em\relax Cambridge University Press, 2003.

\bibitem{ribeiro:10}
A.~Ribeiro, ``{Ergodic Stochastic Optimization Algorithms for Wireless
  Communication and Networking},'' \emph{{IEEE} Trans. Signal Process.},
  vol.~58, no.~12, pp. 6369--6386, Dec. 2010.

\end{thebibliography}


%\end{IEEEbiography}


%\newpage



%\vfill


\end{document}